\documentclass{article}

\usepackage{arxiv}

\usepackage[utf8]{inputenc} 
\usepackage[T1]{fontenc}    
\usepackage{hyperref}       
\usepackage{url}            
\usepackage{booktabs}       
\usepackage{amsfonts}       
\usepackage{nicefrac}       
\usepackage{microtype}      
\usepackage{lipsum}		    
\usepackage{graphicx}
\usepackage{natbib}
\usepackage{doi}
\usepackage{systeme}
\usepackage{amsmath}
\usepackage{comment}
\graphicspath{{Fig/}}

\title{Collective decision making\\ by embodied neural agents}

\author{ \href{https://orcid.org/0000-0000-0000-0000}{\includegraphics[scale=0.06]{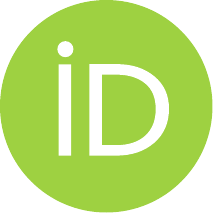}\hspace{1mm}Nicolas~Coucke*} \\
        Consciousness, cognition and computation group\\
        IRIDIA\\
        Université Libre de Bruxelles\\
        \underline{Brussels, Belgium} \\
        PPSP team, CRCHUSJ\\
	Université de Montréal\\
	\underline{Montréal, Canada} \\
        Moral and social brain lab \\
        Universiteit Gent\\
        Ghent, Belgium \\
	\texttt{nicolas.coucke@ugent.be} \\
	\And
	\href{https://orcid.org/0000-0002-1595-8487}{\includegraphics[scale=0.06]{orcid.pdf}\hspace{1mm} Mary~Katherine Heinrich} \\
	IRIDIA\\
	Université Libre de Bruxelles\\
	Brussels, Belgium \\
	\texttt{mary.katherine.heinrich@ulb.be} \\
		\And
	\href{https://orcid.org/0000-0000-0000-0000}{\includegraphics[scale=0.06]{orcid.pdf}\hspace{1mm} Axel Cleeremans} \\
	Consciousness, cognition and computation group\\
	Université Libre de Bruxelles\\
	Brussels, Belgium \\
	\texttt{axel.cleeremans@ulb.be} \\
		\And
	\href{https://orcid.org/0000-0002-3971-0507}{\includegraphics[scale=0.06]{orcid.pdf}\hspace{1mm} Marco Dorigo} \\
	IRIDIA\\
	Université Libre de Bruxelles\\
	Brussels, Belgium \\
	\texttt{mdorigo@ulb.ac.be} \\
		\And
	\href{https://orcid.org/0000-0002-2253-1844}{\includegraphics[scale=0.06]{orcid.pdf}\hspace{1mm} Guillaume Dumas} \\
	PPSP team, CRCHUSJ\\
        Mila - Quebec AI Institute\\
	Université de Montréal\\
	Montréal, Canada \\
	\texttt{guillaume.dumas@umontreal.ca}
}

\date{}

\renewcommand{\headeright}{ }
\renewcommand{\undertitle}{ }
\renewcommand{\shorttitle}{ }
\renewcommand{\shorttitle}{Collective decision making by embodied neural agents}

\hypersetup{
pdftitle={Collective decision making by embodied neural agents},
pdfsubject={q-bio.NC, q-bio.QM},
pdfauthor={Nicolas~Coucke, Mary~Katherine~Heinrich, Axel~Cleeremans, Marco~Dorigo, Guillaume~Dumas},
pdfkeywords={Social NeuroAI, Embodied cognition, Enactivism, Neurodynamics, Multi-agent system, Collective decision making, Swarm Intelligence},
}

\begin{document}
\maketitle

\begin{abstract}
Collective decision making using simple social interactions has been studied in many types of multi-agent systems, including robot swarms and human social networks. However, existing multi-agent studies have rarely modeled the neural dynamics that underlie sensorimotor coordination in embodied biological agents.
%
In this study, we investigated collective decisions that resulted from sensorimotor coordination among agents with simple neural dynamics. 
We equipped our agents with a model of minimal neural dynamics based on the \textit{coordination dynamics} framework, and
embedded them in an environment with a stimulus gradient. 
%
In our single-agent setup, the decision between two stimulus sources depends solely on the coordination of the agent's neural dynamics with its environment.
In our multi-agent setup, that same decision also depends on the sensorimotor coordination between agents, via their simple social interactions.
Our results show that the success of collective decisions depended on a balance of intra-agent, inter-agent, and agent--environment coupling, and we use these results to identify the influences of environmental factors on decision difficulty.
More generally, our results demonstrate the impact of intra- and inter-brain coordination dynamics on collective behavior, can contribute to existing knowledge on the functional role of inter-agent synchrony, and are relevant to ongoing developments in neuro-AI and self-organized multi-agent systems.
\end{abstract}

\textbf{*Corresponding author:} \\ Nicolas Coucke, Department of Experimental Psychology, Henri Dunantlaan 2, 9000 Gent, Belgium, \texttt{nicolas.coucke@ugent.be}

\textbf{Classification} \\
Physical Sciences: Biophysics and Computational Biology \\
Biological Sciences: Psychology and Cognitive Sciences

\section*{Significance statement}
Collective behaviors require the spatial and temporal coordination of actions by  many individuals. The neural mechanisms that enable such coordination among embodied biological agents are currently not well understood. By using simulations of simple embodied agents equipped with biologically plausible neural dynamics, we demonstrated how collective decision making can result from adaptive coupling between an agent's neural dynamics, its environment, and other agents. Our findings make the case for the inclusion of intrinsic neural dynamics in the development of artificial intelligence and multi-agent systems, as a means to expand their ability for social interactions and collective tasks.

\section{Introduction}
Collective decision making is important to the normal functioning of human and animal groups~\citep{Bang2017, Conradt2008} and is also used in groups of artificial agents such as robots~\citep{Hamann2010, MonFerSch-etal2011:si, Valentini2017}.  Decisions can refer to physical actions, such as the direction of movement of animal groups~\citep{Couzin2011}, or symbolic questions that are disconnected from physical situations, such as those studied in collective estimation tasks in humans~\citep{Becker2017, centola2022network}. Collective decisions that are made by the group itself without external intervention typically require that a consenus emerge in the group. Consensus entails that all or at least a large majority of individuals agree, either on an approximate continuous value (e.g., a position in continuous space~\citep{Couzin2005, Yoo2021}) or a discrete option (e.g., voting for an arbitrary item from a list~\citep{Suzuki2015}). 

Consensus is achieved through a distributed process that is not under the control of any single agent~\citep{Valentini2017}. Multi-agent models have been instrumental in investigating how the distributed interactions of individuals can result in a consensus. The individual agents used in most models behave according to rather simple rules or heuristics. For example, in the well-known `opinion dynamics' models, agents typically update their opinion according to the majority or to the voter rule~\citep{Hegselmann2002, Galam2008, Flache2017}. Some recent models aim to replicate human cognitive processes more closely by using neuro-inspired approaches such as the drift-diffusion model~\citep{Srivastava2014, Tump2020, Lokesh2022, Reina2023}. These opinion dynamics models have greatly advanced our understanding of how peer-to-peer interactions can give rise to collective phenomena such as polarization or consensus. 

Another class of multi-agent models addresses collective dynamics in physical space rather than opinion space~\citep{Vicsek1995}. Such models pertain to, for example, animals choosing a new nest site or humans finding an exit during an emergency evacuation~\citep{helbing1995social, Ma2016, reina2017model}. Also, in these more embodied models, behaviors are typically governed by simple rules or heuristics~\citep{Couzin2005, moussaid2011simple}. These models have illuminated, for example, how animals can resolve differences in initial movement directions and move to a single location using only local, implicit communication.

The simple behavioral rules that agents use in these models are approximations of more elaborate brain processes that underlie behavior in biological organisms. Thanks to recent advances in multi-brain neuroscience, it is now possible to simultaneously measure the brain activity of multiple interacting agents during collective behaviors~\citep{kingsbury2020multi}. Despite these innovations, it remains unclear how the combined neural activity of multiple agents is involved in producing collective behavior. Computational models can be of help, by simulating how certain patterns of neural activity across agents produce collective behaviors~\citep{moreau2021beyond}. However, current multi-brain models typically do not link the neural activity of agents to their behavior in an environment; so far, only a handful of multi-agent models use brain-inspired mechanisms rather than simple heuristics to generate collective movements~\citep{resendiz2021shrunken, Sridhar2021, heins2024collective}. In this study, we propose a multi-agent model of collective decision making by embodied agents that are controlled by an oscillatory model of brain dynamics. Our goal with this approach is to pave the way for computational approaches that bridge neuroscience and the burgeoning field of collective behavior.

Considerations of the brain--body--environment interplay have gradually permeated cognitive science, culminating in the \textit{4E cognition framework}, which sees cognitive processes as being embodied, enactive, embedded, and extended~\citep{Clark1998, Newen2018}. Similar considerations have become increasingly common in computational neuroscience and artificial intelligence research~\citep{Steels2018, Colas2022}. 
Largely inspired by the enactive approach to embodied cognition, we constructed minimalist agents that simulate important attributes of biological agents: (1) intrinsic neural dynamics produced autonomously by the agent and (2) a constant sensorimotor loop with the environment~\citep{Varela1992, Thompson2001, Froese2011}. We studied how such agents can reach a consensus by continuously adjusting to one another’s movements in a simple environment. We are not proposing that other approaches to modeling collective decision making are invalid. Rather, we wish to complement prior approaches, by providing a way of incorporating important and understudied aspects of embodied cognitive processes.

In order to successfully operate in an environment, the neural activity of a biological agent must be attuned to the characteristics of that environment. In neuroscience, brain activity is typically studied in terms of oscillations. Neural oscillations (also referred to as brain rhythms) have been linked to perception, movement, and even abstract cognition~\citep{ward2003synchronous, Buzsaki2019}. Across taxa, evolution has selected a subset of rhythms that allow organisms to adequately interact with their environment~\citep{Buzsaki2013}. Moreover, brain rhythms can rapidly shift to accommodate changing environments and task demands~\citep{senoussi2022theta, charalambous2023natural}. 

Large-scale brain rhythms are produced by the coordinated oscillatory activity of many interacting brain regions. The \textit{coordination dynamics} framework has been widely used to study how the activity of these dynamically interacting components is coordinated~\citep{Kelso1997, Kelso2014, Tognoli2020}. One important advantage of the coordination dynamics approach is that it can be used to study the metastable regime in which the brain usually operates~\citep{Tognoli2014}. If coordination among brain regions were always stable, its dynamics could not be adequately modulated by the environment so as to allow the agent to engage in any adaptive behavior. On the other hand, overly unstable dynamics would result in the brain being too easily overwhelmed by environmental input, again preventing adaptive behavior. A metastable regime resolves this problem by allowing the brain to dynamically switch between several stable oscillatory states, thereby being neither completely stable nor completely unstable.

The Haken-Kelso-Bunz (HKB) equations provide a straightforward way to model metastable dynamics among interacting components, such as two populations of oscillating neurons in the brain~\citep{Haken1985, Kelso2014}. Two oscillating components, when modeled with the HKB equations, show in-phase attraction (similar to the Kuramoto model~\citep{Kuramoto1984}) and also anti-phase attraction. The simultaneous existence of in-phase (symmetrical) and anti-phase (asymmetrical) attraction produces a simple form of metastability~\citep{Kelso2013}. The HKB equations were first implemented as a neural controller of an embodied agent by \cite{Aguilera2013}, to model the sensorimotor interactions of a situated agent with its environment. The authors illustrated the importance of taking into account embodied interactions when studying brain dynamics, by showing that the simulated neural dynamics of the agent were qualitatively different when sensorimotor interactions with the environment were disrupted. In this study, we adopted the HKB agent of \cite{Aguilera2013} and modified it so that it could support embodied collective decision making. By making our agent sensitive to both the environment and other agents, we could study the oscillatory neural dynamics of agents when coordinating with both each other and the environment.

In studies of continuous collective decision making, an often-studied question is: under which conditions can agents with different preferred movement directions reach a consensus? For example, two influential modeling approaches have shown that reaching a consensus is facilitated by the presence of a subgroup of unopinionated individuals~\citep{Couzin2005, Leonard2011}. 
In this study, we investigated this ability to reach consensus is modulated by the agents’ neural dynamics. In line with the enactive approach to embodied cognition, we expected that an agent can operate in its environment when it balances (a) its intrinsic neural activity with (b) its sensorimotor coordination with the environment. To investigate this, we first assessed how a single agent’s movement towards one of two local optima in its environment depended on the relative strength of the internal coupling of its brain oscillators versus its coupling to the environment. Agents involved in collective decisions must additionally balance interaction with other agents. Therefore, when we embedded a group of agents with varying initial states in an environment, we assessed how the success of collective decisions depended on a balance of not only intra-agent and agent--environment coupling but also inter-agent coupling (i.e., social influence). Lastly, we connected our model to previous models of collective decision making, by assessing how consensus is influenced by differences in (a) the quality of the two stimulus sources in the environment and (b) differences in individuals’ initial states.

\begin{figure}
\centering
\includegraphics[width=\linewidth]{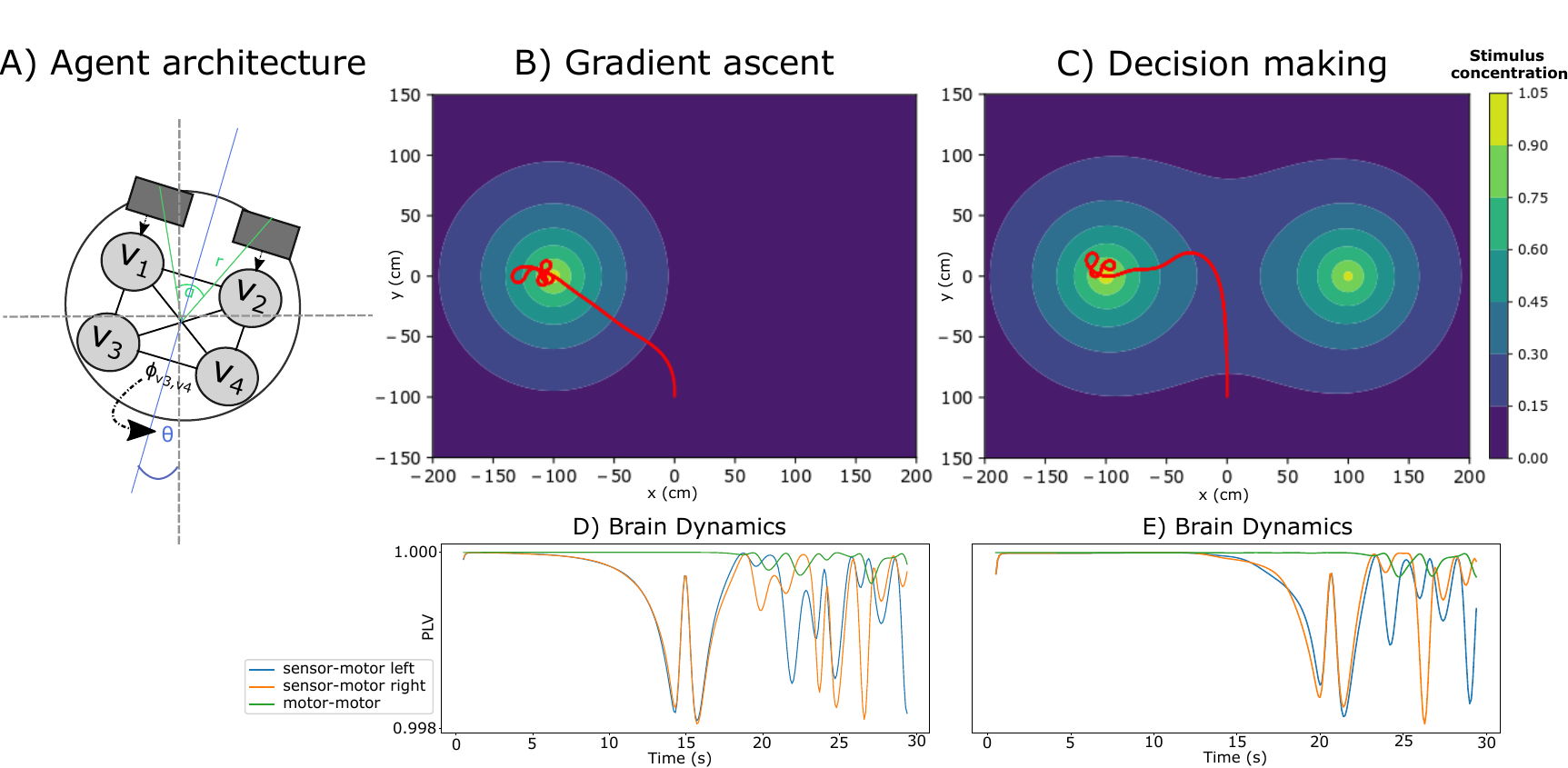} 
\caption{Single-agent behavior and neural dynamics. (A) The agent architecture: two sensors, each connected to a sensory oscillator, nodes 1 and 2 ($v_1$ and $v_2$), that are each in turn connected to a motor oscillator, nodes 3 and 4 ($v_3$ and $v_4$). The traveling orientation $\theta$  of the agent is determined by the angle difference $\phi_{v_3, v_4}$ between motor oscillators. (B) Gradient ascent: the agent's trajectory (red) in the environment (brighter colors indicate higher stimulus concentration). (C) Decision making: two stimulus sources are present in the environment and the agent's performance is measured by its ability to approach one of the two. (D) Internal phase locking of oscillators (contralateral sensor--motor and motor--motor) in B. (E) Internal phase locking of oscillators in C.}
\label{figure1}
\end{figure}

\section{Model}
\label{ch:agent_design}

\subsection{Task environment}
We created a simple environment in which agents could move and sense a stimulus. The environment contained one or more stimulus sources (i.e., sites), at which stimulus concentration is maximal (Fig.~\ref{figure1}B-C). The stimulus concentration in the environment was inversely proportional to the distance from the stimulus source. Thus, the stimulus concentration followed a gradient from low to high concentration when approaching the stimulus source. 
A simple task that agents could perform in this environment was that of gradient ascent, i.e., following the gradient towards maximal stimulus concentration (Fig.~\ref{figure1}B)~\cite{Aguilera2013}. If one imagines the stimulus as ‘food,’ and the stimulus source as a food source, then this behavior reflects the food-seeking behavior of many simple organisms. 
When two sources of stimulus are present, the scenario could be considered a binary decision-making task (Fig.~\ref{figure1}C). In this scenario, an agent could successfully reach a stimulus source if it could ‘decide’ between the two sources. 
In the multi-agent scenario, the task became a collective binary decision-making task. In this scenario, 10 agents started at the same position, but had different initial orientations (Fig.~\ref{figure2}B). Due to these different initial orientations, agents could end up at different sites (Fig.~\ref{figure2}C). However, agents had some social information about each other’s position (Fig.~\ref{figure2}A; see below), and their task was to use this information to aggregate at the same site. We quantified performance of collective decision making according to how closely the agents collectively approached a single candidate site (see Methods).

\subsection{Agent}
We modeled an agent with minimal neural dynamics that could use sensorimotor coordination with the stimulus sources and the movements of other agents in order to move towards a candidate site. Our agent architecture was based on a minimal Braitenberg vehicle~\citep{Braitenberg1986}, which is a self-driven agent with a very simple architecture: two sensors directly control two motors. 
To give our agent intrinsic neural dynamics, we connected two oscillator nodes to the sensors (loosely representing sensory brain regions; nodes 1 and 2 in Fig.~\ref{figure1}A) and two oscillator nodes connected to the direction of the movement of the agent (loosely representing motor regions; nodes 3 and 4 in Fig.~\ref{figure1}A). This design resembles the \textit{situated HKB agent} of \cite{Aguilera2013}, which had two oscillator nodes (one sensory and one motor). Our agents have four nodes, so that they can use stereovision and differential drive to move directly to a stimulus source, rather than approaching it in a spiraling motion~\citep[cf.]{Aguilera2013}.

To model the interaction between the oscillators, we used an update rule for the phase of each oscillator, based on the following version of the HKB equation~\citep{Zhang2019}:
\begin{equation}
    \dot{\varphi}_{v_i} = \delta \omega_{v_i} + c I_{v_i} - \sum^{N}_{j=0} a_{v_i, v_j} \sin{\phi_{v_i, v_j}} - 2 b_{v_i, v_j}  \sin{2\phi_{v_i, v_j}}~,
\label{eq:HKB_model}
\end{equation}
where $\dot{\varphi}_i$ is the phase change of node $v_i$, and $\omega_{v_i}$ is the intrinsic frequency of oscillator $v_i$. Parameters $a_{v_i, v_j}$ and $b_{v_i, v_j}$ represent the contribution of, respectively, in-phase attraction, and anti-phase attraction between oscillators $v_i$ and $v_j$. Lastly, $c$ parameterizes how strongly the oscillator phase is modulated by sensory input $I_{v_i}$. This parameter is set to zero for each motor oscillator, because it is not connected to a sensor. (See Methods for the version of the update equation used for each oscillator.) 

Our agent moves at a constant speed and the activity of the motor oscillations is linked to the agent's movement direction. The heading is updated according to the phase angle between the two motor oscillators, such that
\begin{equation}
    \dot{\theta} = \eta \phi_{v_3, v_4}~,
\end{equation}
where $\theta$ is the orientation of the agent in the environment and $\eta$ is a scaling factor. Together, these equations create a closed sensorimotor loop between the agent's internal oscillator dynamics and the external environment.

In the multi-agent scenario, we gave agents the added behavior of emitting the same stimulus that they observed to be present in the environment. The stimulus concentration emitted by social agent $j$ was perceived by agent $i$ as:
\begin{equation}
    I_{ij} = S*e^{-\lambda_a D_{ij}}~,
\label{eq:social_stimulus}
\end{equation}
where $D_{ij}$ is the Euclidean distance between agent $i$ and agent $j$, $S$ is the strength of social influence between agents (identical for all agents), and $\lambda_a$ is the decay rate of the emitted stimulus. Note that an agent did not perceive its own emitted stimulus. 

\section{Results}

We performed both single-agent simulations and multi-agent simulations. For the single-agent simulations, we quantified neural coordination dynamics in terms of integration and metastability. The integration of brain regions by means of phase-locked activity is a central mechanism of brain function ~\citep{Varela2001, avena2018communication}, and can be quantified using the phase-locking value (PLV). Brain function supportive of adaptive behavior relies on switching between different brain states. To quantify this aspect of neural dynamics, we used the standard deviation of the Kuramoto order parameter SD(KOP)~\citep{Strogatz2000, Cabral2022}. 

For the multi-agent simulations, we additionally quantified the coordination dynamics occurring across the different agents. We analyzed agents' movement trajectories using the KOP as a measure of alignment, and SD(KOP) as a measure of alignment variability between agents' movements~\citep{Strogatz2000, Cabral2022}. Lastly, we also quantified the degree of coordinated activity between the neural dynamics of the different agents by using the weighted phase-lag index (wPLI), a measure of phase locking that discards zero-phase coupling and can be interpreted as the co-variance between two signals~\citep{Vinck2011}.

\subsection{Single-agent simulations}

\begin{figure}[h]
\centering
\includegraphics[width=\linewidth]{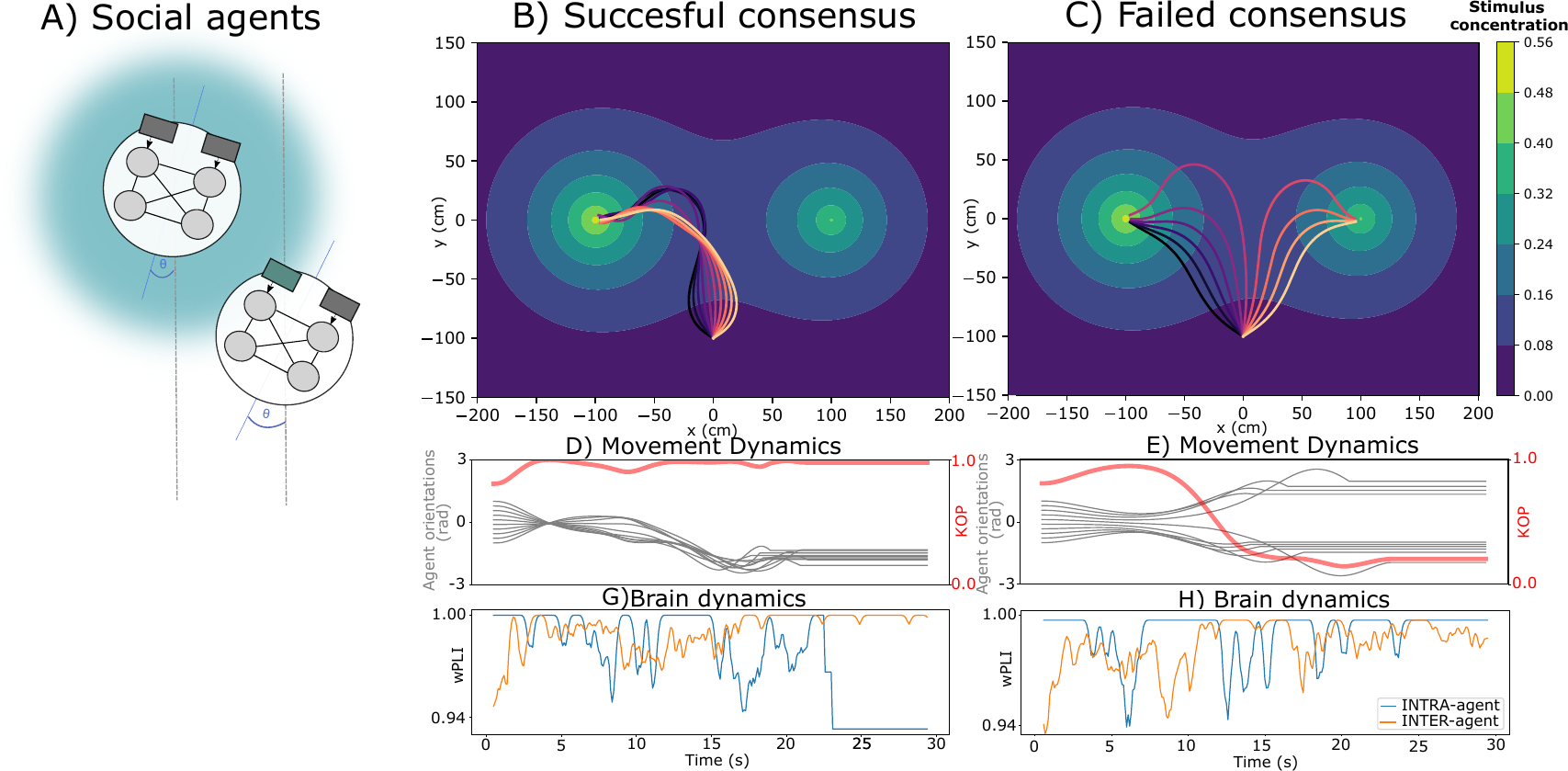} 
\caption{Agent behavior and intra-agent neural dynamics during collective decision making. (A) Agents emit stimulus that can be perceived by other agents. (B) Higher social stimulation allows agents to converge onto the same stimulus source in their environment. (C) With lower social stimulation, agents do not converge on the same source. (D-E) Movement angles of agents (gray lines) and KOP of the group, indicating the degree of alignment (red line). Alignment increases when all agents are moving towards the same stimulus source and decreases when they are not. (G-H) Intra- and inter-agent neural dynamics: average intra-agent wPLI (blue line) and average inter-agent wPLI (orange line).}
\label{figure2}
\end{figure}

\begin{figure}
\centering
\includegraphics[width=\linewidth]{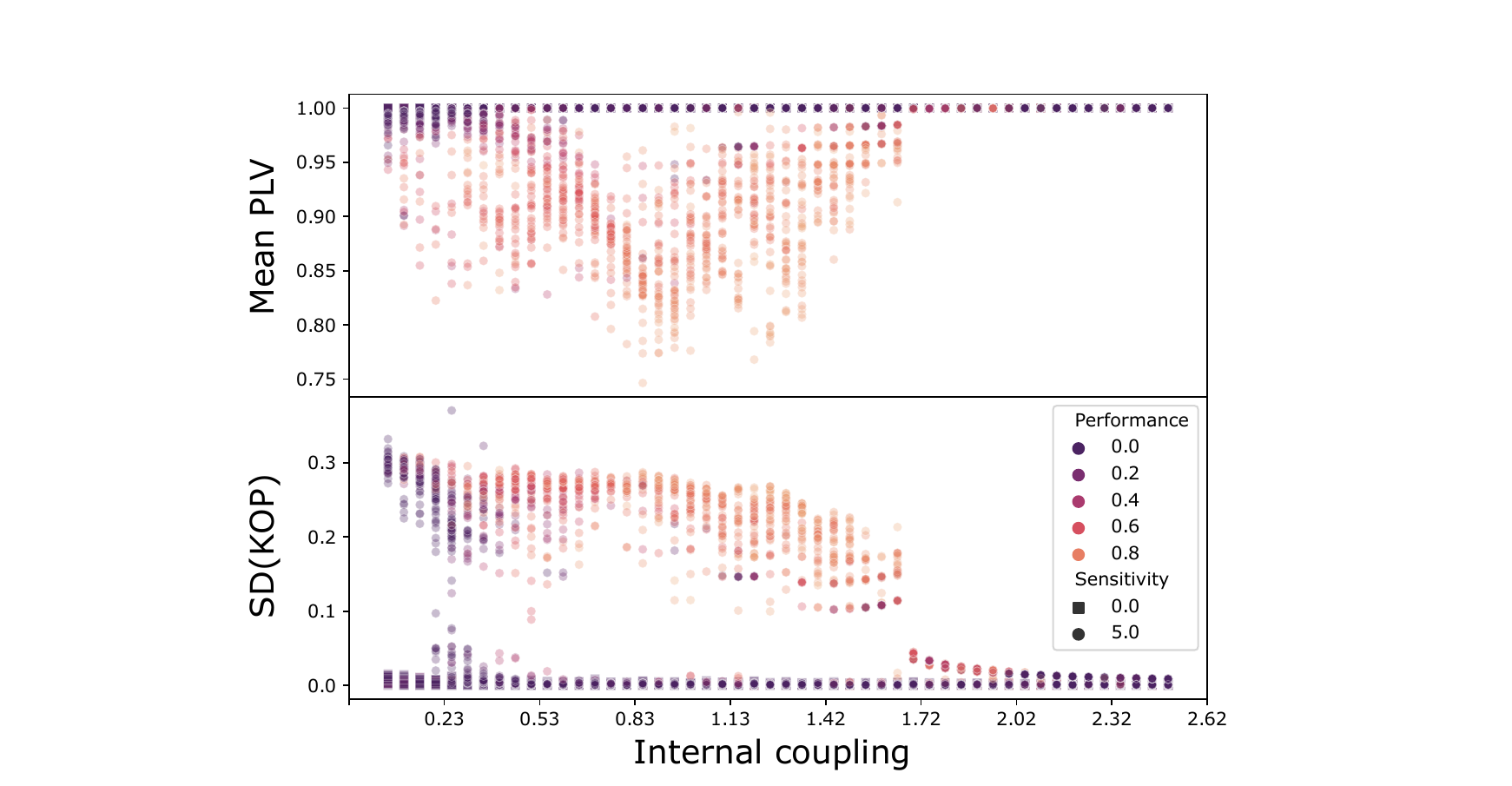} 
\caption{Intra-agent neural dynamics: the mean PLV and SD(KOP) of each agent during its run ($y$ axis), according to the internal coupling degree ($a_{v_i,v_j}$ and $b_{v_i,v_j}$) in its neural controller ($x$ axis). 
Squares represent agents without sensory input and dots represent agents with sensory input. 
Lighter colors represent higher performance.
For each configuration (i.e., internal coupling degree and stimulus sensitivity), 50 runs were performed with random initial phases of the oscillators. Each data point represents the average of one run. 
}
\label{figure3}
\end{figure}

We first performed a series of single-agent simulations to assess how an individual agent's neural dynamics are related to its ability to move towards a stimulus source in its environment. In the single-agent setup (see Fig.~\ref{figure1}), an agent tried to climb a gradient towards a global maximum. To simulate different types of neural dynamics, we varied the internal coupling strength between the agent's oscillator nodes ($a_{v_i,v_j}$ in Eq. \ref{eq:HKB_model}), and varied whether or not it was sensitive to external stimuli. For each agent configuration, we performed 50 runs with different random initial oscillator phases. In Fig.~\ref{figure3}, we characterize the neural dynamics associated with each agent configuration. The top panel shows the average level of integration (i.e., phase locking) between the agent's oscillators, measured by the mean PLV, and the bottom panel shows the degree of metastability among the agent's oscillators (measured by SD(KOP); see Methods). 

It is notable that, in the absence of sensory input, the system quickly found a stable state with minimal variation in oscillator dynamics: agents without stimulus input (squares in Fig.~\ref{figure3}) consistently had values close to PLV$=1$ and SD(KOP)$=0$. Conversely, agents with sensory input (circles in Fig.~\ref{figure3}) had a broad range of parameter values that resulted in lower PLV and higher SD(KOP). This shows that, as expected, sensory input can alter the coordination regime of the neural dynamics. In the presence of sensory input, PLV decreases as internal coupling increases from 0 to 1, indicating that the neural dynamics at low internal coupling are mostly driven by stimulus input, without being significantly modulated by the interactions among the agent's own oscillators. Simultaneously, SD(KOP) remained relatively high, indicating that, at low internal coupling, sensory input caused the system of oscillators to quickly cycle between oscillatory states. 

As the agent's internal coupling increases further, the interactions between the agent's oscillators become strong enough to meaningfully modulate sensory input, which results in a lower level of apparent oscillator integration, with PLV decreasing to 0.75 while the degree of metastability plateaus. Beyond an internal coupling level of $a_{v_i,v_j}=1.4$, the internal coupling of the agent's oscillators started to dominate, which resulted in highly integrated oscillators, indicated by higher PLV. At an internal coupling level of $a_{v_i,v_j}=1.7$, the effect of internal coupling became strong enough that it nullified the effect of any sensory input, resulting in PLV$=1$ (indicating no variation in inter-oscillator dynamics). This increase in integration was accompanied by a sharp drop in metastability, indicating that the system tends to get stuck in a single stable state. Such a stable state of high integration between oscillators precludes changes in movement direction in response to sensory input, inhibiting the agent from approaching the stimulus source.

The colors of the data points in Fig.~\ref{figure3} indicate agent performance: brighter colors indicate that the agent was better able to approach the stimulus. The distribution of colors shows that agents performed best in the range $a_{v_i,v_j} \in \{0.8, \ldots ,1.5\}$. At these intermediate coupling values, there was a decrease in oscillator integration (as shown by PLV) and moderately high metastability (as shown by SD(KOP)). In short, these results show that at low internal coupling, neural dynamics are mostly driven by sensory input, and oscillatory coordination states change quickly. At moderate internal coupling, neural dynamics modulate sensory input without fully dominating it. In this intermediate range with relatively low integration and high metastability, the agent displays adaptive behavior. At high internal coupling, neural dynamics nullify the effect of sensory input and agent behavior cannot change in response to the environment. (See the supplementary text S2 for a more elaborate analysis of gradient climbing and decision making by single agents, as well as its relation to neural dynamics.)

\subsection{Multi-agent simulations}

We ran two sets of multi-agent simulations. In the first set of simulations, we varied parameters related to the agent configuration (internal coupling, environmental sensitivity, social influence) and observed the effects on decision-making performance and collective dynamics. In the second series of experiments, we kept the agent configuration constant and modified the quality difference between the two stimulus sources in their environment, as well as the starting angles between the agents.

\subsubsection{Consensus as a function of internal, environmental, and social influences}

We conducted a series of simulations with groups of 10 agents in an environment with a fixed quality (brightness) ratio of $r = 0.8$ between the two stimulus sources (Fig.~\ref{figure2}). For each simulation run, we quantified the performance as the degree to which agents could approach the same stimulus source (as in Fig.~\ref{figure2}B) rather than going to different sources (as in Fig.~\ref{figure2}C; see Methods). Throughout each simulation, we track the coordination dynamics among agents' movements (Fig.~\ref{figure2}D-E), as well as measures of coordination between the neural dynamics of the different agents (Fig.~\ref{figure2}G-H).

\begin{figure}
\centering
\includegraphics[width=\linewidth]{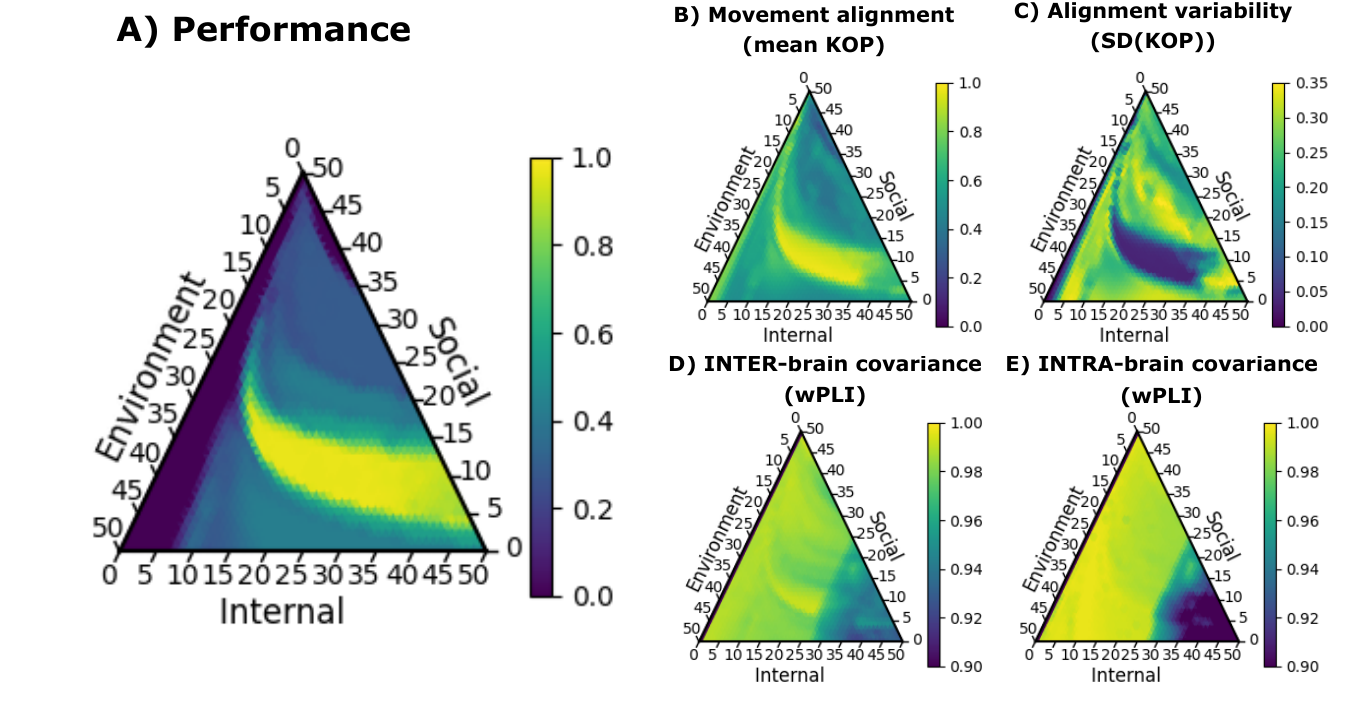} 
\caption{Ternary plots illustrating how collective behavior and neural dynamics depend on the agent configuration. Each point in the triangle corresponds to a certain weighting of environmental stimulus, social information, and internal motor coupling. In each simulation, 
the parameters 
fulfill the condition $stimulus \, sensitivity + social \, sensitivity + internal \, coupling = 100$. The scale $[0,50]$ for each dimension corresponds to respective parameter values  of $c \in \{0, \ldots ,10\}$ for stimulus sensitivity,  $S \in \{0, \ldots ,5\}$ for social sensitivity, and $a_{v_3,v_4} \in \{0, \ldots ,1\}$ for internal coupling. The top corner corresponds to maximal social sensitivity, the left corner to maximal environmental sensitivity, and the right corner to maximal internal coupling. The brightness (yellowness) in panel A indicates the performance of collective decision making. A performance of 0 indicates that agents failed to reach either of the two stimulus sources. A performance of 0.5 indicates that half of the agents reached the same stimulus source. A performance of 1 indicates that all agents reached the same stimulus source and thus that a consensus was reached. The brightness in panels B-E indicate the strength of, respectively, the movement alignment, alignment variability, inter-brain covariance, and intra-brain covariance.
}
\label{figure5}
\end{figure}

We conducted simulations for different parameter values of internal coupling strength ($a_{v_i,v_j}$ in Eq.~\ref{eq:HKB_model}), sensitivity to the environment ($c$ in Eq.~\ref{eq:HKB_model}), and the degree of social influence ($S$ in Eq.~\ref{eq:social_stimulus}). For each combination of parameter values, we display the final performance in a ternary plot (Fig.~\ref{figure5}A). Each of the three corners of the ternary plot corresponds to one of the parameters being maximal and the others zero. We also display the corresponding measures of movement coordination and neural coordination dynamics for each parameter combination in adjacent plots (Fig.~\ref{figure5}B-E).

We assessed movement coordination in terms of the movement alignment (KOP) and the variability in movement alignment (i.e., 'alignment variability'; SD(KOP)). To assess the coordination dynamics within and between agents’ neural dynamics, we used a measure of phase covariance (wPLI), rather than the phase-locking value used for the single-agent case. In contrast to the PLV, the wPLI does not take into account zero-lag coupling between oscillators. Measures with this property are preferred in multi-brain neuroscience, since they discount spurious coordination due to, e.g., common input from the environment~\citep{Czeszumski2020, Schwartz2022}. In our simulations, spurious coordination between oscillators could similarly have been caused by identical initial phases of agents' oscillators (see Methods).

The middle region of the plots in Fig.~\ref{figure5} corresponds to a parameter range in which internal, environmental, and social influences are appropriately  balanced for reaching consensus, as indicated by the bright yellow area in Fig.~\ref{figure5}A. This region was accompanied by high movement alignment and low  alignment variability, indicating that agents could use environmental and social information to coherently move towards the same stimulus source. Part of this region corresponds to a narrow area of increased inter-brain covariance, indicating that this aligned movement was accompanied by coordinated neural dynamics across agents. 
The lower right corner of the plots corresponds to a region of increased internal influences and decreased social influences. As long as social influences are non-zero, performance remains relatively high in this parameter range. Movement alignment is decreased and alignment variability is increased relative to the middle part of the plot. This indicates that, when internal coupling increases, agents’ movements become less aligned, but can still result in consensus. Interestingly, the lower right corner of the plot corresponds to a decrease in both intra-brain covariance (between different oscillator nodes within the same agent) and inter-brain covariance (between the same oscillator nodes across different agents). More alignment variability among the agents' movements is thus accompanied by more independent by brain dynamics that are more independent.

A last and interesting observation can be made at the left edge of the plots, where internal coupling is low. In this region, agents are driven entirely by a combination of social and environmental influences. This region in parameter space was accompanied by high movement alignment and low alignment variability, but was associated with a decrease in performance. This suggests that agents moved in a highly aligned manner, but failed to collectively approach either of the two stimulus sources. This outcome highlights the importance of balancing external influences with sufficient internal coupling. When agents are overly coupled to external stimuli without enough counteracting internal coupling, they struggle to move towards an increasing stimulus concentration. Supplementary figure~S2 illustrates how increased social influence, without a corresponding increase in internal coupling, leads to decreased performance.

Taken together, these results show that agents reach a consensus when their configuration facilitates a balanced integration of environmental, social, and internal influences, and this is reflected in neural and behavioral dynamics.

\subsubsection{Consensus as a function of environment configuration}
If the ability to reach a consensus depended on the features of the environment, we would expect that binary decision making should be easier (and thus performance higher) when the difference between the two stimulus sources is larger, and when the initial angle between the agents is smaller. We performed simulations with 10 agents with a fixed architecture, and varied the initial starting angle between agents and the brightness ratio between the two stimulus sources (see Methods).

\begin{figure}
\centering
\includegraphics[width=\linewidth]{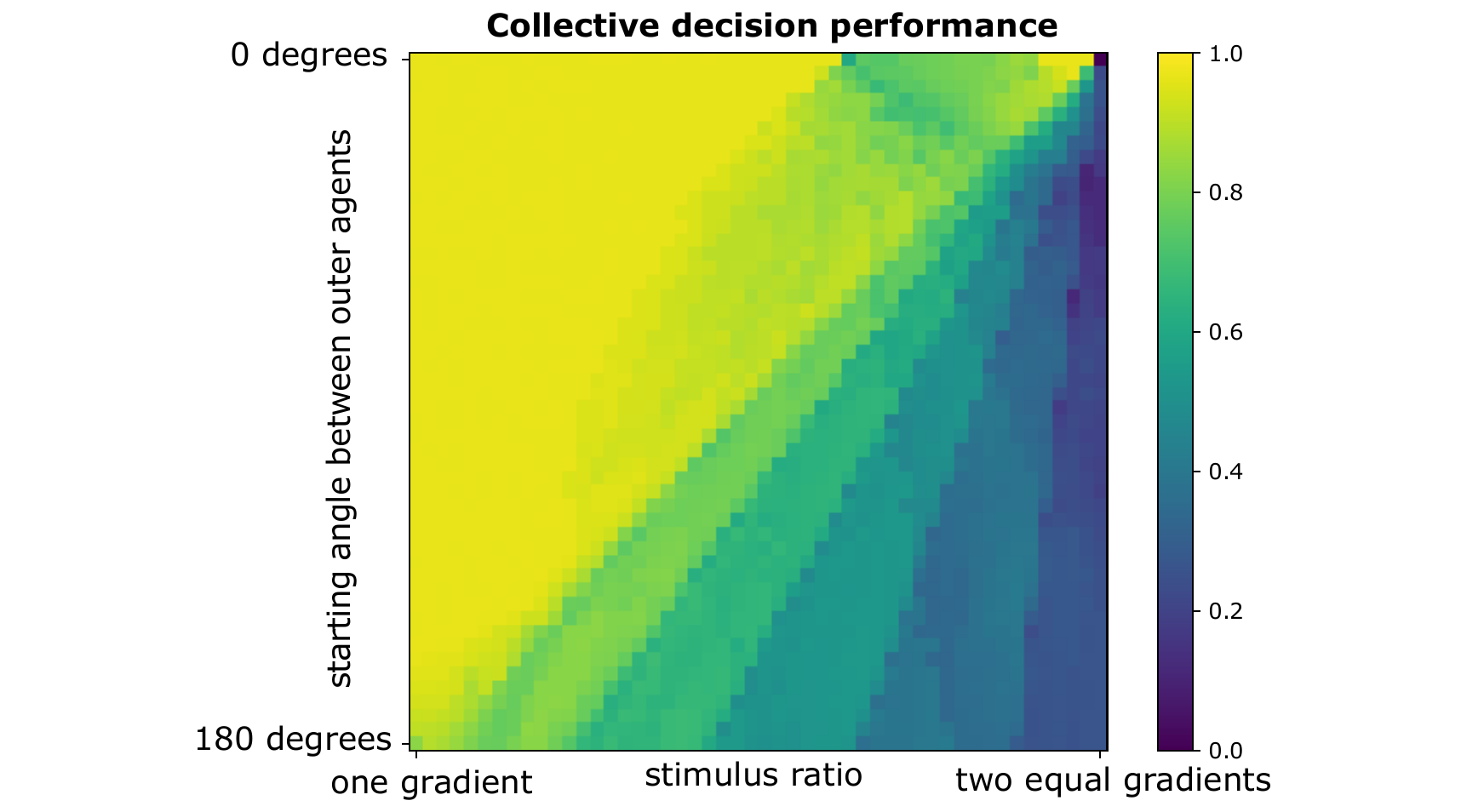} 
\caption{Dependence of the collective decision-making performance on the environment and initial orientations of agents. The leftmost extreme of the $x$-axis represents the cases with only one stimulus source present in the environment. Moving towards the right, the brightness of a second stimulus source increases until the two have equal brightness. The agents always start with equal angles between them. At the bottom of the $y$-axis, the agents are spread so that the outermost two of the ten agents are at a 180° angle. At the top of the $y$-axis, all agents start with angles of 0° between them. All agents have identical parameters; stimulus sensitivity is $c=3$, social sensitivity is $S=1$, and internal coupling is $a_{v_i,v_j}=0.5$.}
\label{figure7}
\end{figure}

The results in Fig.~\ref{figure7} show that, overall, performance depended on a combination of starting angle and stimulus ratio. Performance was maximal when agents had identical starting orientations and only one stimulus source was present (top left of Fig.~\ref{figure7}). In accordance with our expectations, performance decreases as the second stimulus source became brighter and the starting angle between agents increased. 

It should be noted that performance did not increase linearly as the decision-making task became easier. Rather, there was a repeating pattern of sharp decreases in performance followed by short plateaus. This was most likely due to a combination of our performance measure and the relatively small number of agents (see Methods). Each drop in performance was caused by one of the ten agents moving away from the global maximum and towards the competing local maximum (i.e., to the stimulus source with lower brightness). Supplementary figure~S3 provides a more detailed account of this pattern. Overall, these simulations show that the ease with which the agents reach a consensus depends not only on their architecture but also on the environment in which they operate.

\section{Discussion}

We have modeled collective decision making with embodied neural agents that are controlled by simple oscillatory brain dynamics. We first showed that the ability of an agent to move towards a stimulus source was reflected in the intra-brain dynamics of that agent. In an intermediate parameter range where agents were neither overwhelmed by environmental stimulus nor insensitive to it, neural dynamics could become sufficiently uncoupled and metastable to allow the agent to approach its target. Furthermore, multiple agents with different initial heading directions could overcome their differences and converge on one of the two available options. Agents were able to do this within a parameter range that adequately weighted environmental, social, and internal influences. When one influence was too high with respect to the others, the agents failed to converge on a decision. 

In this regard, our model differs from more disembodied multi-agent models~\citep{Flache2017}. In disembodied cognition models of collective decision making, increasing social influence to an arbitrarily high degree would lead to a fast and efficient convergence on any of the possible decision options. The option selected may be of greater or lesser quality, but the ability of agents to reach any option at all has rarely been studied. Our results showed that, when increasing social sensitivity too much, the agent's neural dynamics may become saturated with social information---to such a degree that it cannot adequately interact with the environment and move towards one of the options. This is reminiscent of real-world situations in which agents that are too consumed by interacting socially with one another lose adaptive interactions with the environment (see Strasbourg dancing plague~\citep{Waller2008} or circular milling in army ants~\citep{Couzin2003a}).


Our model also differs from the more embodied multi-agent models of movement-based collective decisions by animals. In most such models, agents have a parameter that explains their preferred target location or movement direction~\citep{Couzin2011, Leonard2011, Sridhar2021}. In our simulations, agents had different initial movement directions but did not have a parameter representing a preferred movement direction. Rather, their movement direction emerged from their interactions with each other and with their environment. Our results showed that the more an agent's initial movement direction was between the two sources rather than pointing to one of them, the more likely agents were to reach a consensus. These results are somewhat in line with previous findings that a larger proportion of unopinionated individuals promotes consensus in models of movement-based decisions made by groups of animals?)~\citep{Couzin2011, Leonard2011}.

Our simulations also showed that a larger difference between the quality of stimulus sources in the environment generally resulted in a higher proportion of agents reaching a consensus. This is in accordance with studies of discrete decision making between a few options in an environment, where models have shown how the speed and accuracy of collective decisions depend on differences between environmental stimuli~\citep{Valentini2015, Valentini2017}. Models that do not take into account the quality of options in the environment, or that consider options of equal quality, often observe agents traveling in a compromise direction~\citep{Leonard2011}. Since agents' neural dynamics were strongly influenced by the environment in our simulations, such compromises only occur in a small region of the parameter space. The closer agents came to a stimulus source, the higher the stimulus concentration they observed, causing agents to almost always go to one of the two stimulus sources instead of taking a compromise direction. Our simulations also showed some surprising emergent collective behaviors, such as `overshooting` in response to social influence and, as a result, moving towards an option that corresponds neither with the initial movement direction nor with the option chosen by other agents (see Supplementary figure~S2A).

In the current paper, we performed deterministic simulations of agents that started from the same position and moved towards one of two stimulus sources. In future work, our model could be studied in environments with a higher number of stimulus sources, without changes to the agent architecture~\citep{Sridhar2021}. Furthermore, agents might be allowed to visit multiple options of stimulus sources before converging on one, as is common in collective decisions of ants and honeybees~\citep{Hoelldobler1990, reina2017model}. Another extension of our simulations could be to let agents start from different spatial locations. Using the HKB equations to maintain asymmetric patterns of coordination between different agents' oscillators, future work could study a rudimentary allocentric way of using social information~\citep{Pickavance2018}. Future work could also study the influence of noise on collective performance of embodied neural agents, especially in environments with many local optima, as random fluctuations are often an important aspect of self-organized systems and collective intelligence~\citep{Bonabeau1999, Kahneman2022}. 

Inspired by the brain dynamics of biological agents, we used oscillator models to study neural dynamics in a multi-agent system. Recent \textit{swarmalator} models have also combined collective movement with oscillator dynamics~\citep{OKeeffe2017, Ceron2023}. In these models, an agent's behavior is based on the directly observed oscillator phases of the surrounding agents. In our model, the oscillators represent brain dynamics that are not directly available to an outside observer. The brain dynamics of the agents can only become coordinated by intermediary of their behavior. Moreover, since the stimulus emitted by agents was indistinguishable from stimulus originating in the environment, agents could not selectively react to instantaneous social stimulation. Yet, agents could interact by mutually reacting to local changes in stimulus concentration caused by their respective movements in the environment. This situation reflects a form of social interaction in which agents cannot use any social cognition capabilities other than those involved in the interaction itself. Such situations are also studied with human participants in perceptual crossing experiments, and have led some to argue that social interaction can be constitutive of social cognition~\citep{auvray2009perceptual, DeJaegher2010}.

Using oscillators to model brain dynamics allows us to use phase-locking and phase-covariance measures to quantify the degree of coordination between the brain dynamics of different agents~\citep{Czeszumski2020}. Experimental studies with multiple animals or humans have suggested that coordination between brain dynamics of different agents, typically quantified in terms of inter-brain synchronization (IBS), might have an important role in supporting collective behaviors~\citep{Dumas2010, Yang2021}. A few computational models have provided initial mechanistic explanations for the emergence of such inter-brain synchrony by using the Kuramoto model, showing that the strength and frequencies at which IBS takes place depend on a combination of agents' individual brain dynamics and their inter-agent coupling~\citep{Dumas2012a, Moreau2022}. Although our results cannot conclusively show whether collective decision-making performance depended on inter-agent synchrony, our models could provide a way to study the complex brain--brain behavior dynamics that can give rise to IBS. Furthermore, our results replicate an interesting finding of Kuramoto models of interpersonal synchrony, namely that some degree of intra-agent coupling is required to achieve rich patterns of interpersonal coordination~\citep{Heggli2019}. While previous studies (e.g.,~\citep{Heggli2019}) have shown this requirement when a pair of agents were coupled to each other directly, we have confirmed it for multiple agents embedded in a spatial environment. 

A major challenge in the development of artificial agents is coordinating social interactions with both humans and other artificial agents. Recent developments in Social NeuroAI attempt to bring social interaction into the realm of AI by advancing artificial agents' social embodiment, temporal dynamics, and biological plausibility~\citep{Bolotta2021}. In this work, we accommodate 1) social embodiment, as collective decisions are movement-based and agents can be reciprocally influenced by each other’s movements; 2) temporal dynamics, through continuous intra-agent, inter-agent, and agent--environment interactions; and 3) biological plausibility, by using oscillations to control agent behavior. Our approach could be a starting point for developing social-neural agents that collaborate on a wider range of collective tasks through the implicit coordination of their neural dynamics.

\section{Methods}

\subsection{Experiment setup}


Each experiment took place in a 2D environment in which every position had an associated stimulus concentration. Depending on the experiment type, each environment contained one or more stimulus sources of different quality. 
The stimulus concentration at a certain position was exponentially proportional to its closeness to the stimulus source: 
\begin{equation}
    I(x,y) = e^{-\lambda D(x,y)},
\label{eq:env-stimulus}
\end{equation}
where $D(x,y)$ is the Euclidean distance to the stimulus source and $\lambda = 0.02$ is the exponential decay rate of the environmental stimulus.
In setups with two stimulus sources, each had stimulus concentrations defined by eq.~\ref{eq:env-stimulus}, and the overall stimulus concentration at a certain position was the combination of the two:
\begin{equation}
    I(x,y) = e^{-\lambda D_{1}(x,y)} + r e^{-\lambda D_{2}(x,y)}~,
\label{eq:env-stimulus-2-sources}
\end{equation}
where $r$ indicates the quality ratio of the two stimulus sources.
When there were multiple agents in an environment, the stimulus level that an agent perceived was a combination of the stimulus concentration in the environment and the stimuli concentrations emitted by other agents, such that
\begin{equation}
    I_{i}(x,y) = e^{-\lambda D_{1}(x,y)} + r e^{-\lambda D_{2}(x,y)} + S*\sum_{j}{e^{-\lambda_s D_{ij}}}~.
\label{eq:two_stimuli}
\end{equation}


In all experiments, the environment was 300 by 400 cm, the radius of each agent's body was 2.5 cm, and each agent has a fixed velocity of 10\,cm/s.
Simulations were performed with a timestep of 0.01\,s. All experiments ended after 30\,s, which provided sufficient time for agents to reach a stimulus source in the environment. 
All simulations were performed in Python version 3.9.2~\citep{Rossum} and the agents were implemented in Pytorch version 1.12.0~\citep{PyTorch2019}. 
The code is available in an open-source code repository: \url{https://github.com/ppsp-team/PyHKBs}.

\subsection{Agent design}

In our agents, sensory input did not directly control motor activity. Sensory information (in the form of stimulus concentration) was first integrated into the oscillator phase of two sensory nodes. These sensory nodes were dynamically connected to two motor nodes. 

The situated agent designed by \cite{Aguilera2013} consisted of one motor oscillator and one sensory oscillator, and thus could only perform gradient ascent with spiraling movement. We resolved this by giving our agent two sensory oscillators for stereovision ($v_1$ and $v_2$, see nodes 1 and 2 in Fig.~\ref{figure1}A) and two motor oscillators for differential drive steering ($v_3$ and $v_4$, see nodes 3 and 4 in Fig.~\ref{figure1}A).
The sensors are directionless and are placed at the front of the agent, 90° apart as measured from the agent's center (see Fig.~\ref{figure1}a).
The orientation $\theta$ of the agent in the environment is determined by the angle between the two motor oscillators ($v_3$ and $v_4$, see nodes 3 and 4 in Fig.~\ref{figure1}A).

Altogether, the dynamics of the agent are governed by the following set of equations:

\begin{equation} 
\dot{\boldsymbol\varphi} 
\begin{cases}
    {\dot{\varphi}_{v_1}} = {\omega_{v_1} + c I_l - \sum_{j} a_{v_1,v_j} \sin{\phi_{v_1,v_j}} - \sum_{j} b_{v_1,v_j} \sin{2\phi_{v_1,v_j}}},\\
    {\dot{\varphi}_{v_2}} = {\omega_{v_2} + c I_r -\sum_{j} a_{v_2,v_j} \sin{\phi_{v_2,v_j}} - \sum_{j} b_{v_2,v_j} \sin{2\phi_{v_2,v_j}}},\\
    {\dot{\varphi}_{v_3}} = \omega_{v_3} - \sum_{j} a_{v_3,v_j} \sin{\phi_{v_3,v_j}} - \sum_{j} b_{v_3,v_j} \sin{2\phi_{v_3,v_j}},\\
    {\dot{\varphi}_{v_4}} = \omega_{v_4} - \sum_{j} a_{v_4,v_j} \sin{\phi_{v_4,v_j}} - \sum_{j} b_{v_4,v_j} \sin{2\phi_{v_4,v_j}}\\
\end{cases},
\label{eq:HKBsocial}
\end{equation}
\begin{equation}
\dot{\theta} = \eta\phi_{v_3,v_4} = \eta(\varphi_{v_3} - \varphi_{v_4})~,
\end{equation} 
where we fixed the ratio $k = {\frac{b}{a}} = 2$ so that the HKB equations are bistable (see supplementary text S1).

In our neural controller, the oscillators influenced each other over the following connections: the contralateral ones ($a_{v_1,v_4}=a_{v_4,v_1}$ and $a_{v_2,v_3}=a_{v_3,v_2}$) the one between the motor regions ($a_{v_3,v_4}=a_{v_4,v_3}$), as well as their antiphase counterparts.
Thus, we kept the two sensory oscillators independent and incorporated the contralateral sensorimotor connections present in the Braitenberg vehicles~\citep{Braitenberg1986} and in many of the biological neural organizations~\citep{Sterling2017}.
The intrinsic frequencies of all oscillators were set to 5\,Hz, to resemble the frequency of the theta oscillations in biological brains.
In our model, the next phase of each oscillator is calculated at each time step, by integrating the differential equations using the fourth-order Runge-Kutta method.




\subsubsection{Single-agent experiments}

In the single-agent gradient ascent setup,
the agent initiated movement at $xy$ position (0, -100) and the stimulus source was located at $xy$ position (-100, 0).
To study the link between a single agent's behavior and its intra-agent neural dynamics, we varied the sensory sensitivity ($c=$ 0 or 5, in Eq.~\ref{eq:HKBsocial}) and the coupling strength of all connections ($a_{v_i,v_j}$ values from 0.05 to 2.5, in steps of 0.05, in Eq.~\ref{eq:HKBsocial}). 
For each variation combination, we performed 50 runs with random initial phases of the oscillators.

In the single-agent binary decision making setup, the first stimulus source was located at $xy$ position (-100, 0), the second at $xy$ position (100, 0), and the brightness (i.e., quality) ratio of the two stimulus sources is $r = 0.95$ (see Eq.~\ref{eq:two_stimuli}). 
The agent initiated movement equidistant to the two stimulus sources, at $xy$ position (0, -100) with all internal oscillators starting as in-phase.
To evaluate the dependence of performance on agent behavior, we varied the stimulus sensitivity of the agent ($c$ values from 0 to 10, in steps of 1, in Eq.~\ref{eq:HKBsocial}) and the internal coupling ($a_{v_i,v_j}$ values from 0.05 to 2.5, in steps of 0.05, in Eq.~\ref{eq:HKBsocial}). To evaluate the importance of internal coupling, we also varied whether the motor regions were connected or not ($a_{v_3,v_4}=0$, in Eq~\ref{eq:HKBsocial}). We ran one simulation for each variation combination, since we did not introduce an element of randomness in the simulation.

\subsubsection{Multi-agent experiments}

Each multi-agent experiment had a group of 10 agents and an environment with two stimulus sources, located at $xy$ positions (-100, 0) and (100, 0).
To study consensus achievement under divergent starting opinions, we evenly distributed the initial orientations of the agents (between angle $-\theta_{\texttt{max}}$ and $\theta_{\texttt{max}}$, $\theta_{\texttt{max}} \in [0^{\circ}, 180^{\circ}]$). Thus, each agent faced a different initial direction, with half facing more towards the lefthand stimulus source and half facing more towards the righthand one. 
Following \cite{Nabet2009, Leonard2011}, the agent behavior in these experiments was deliberately deterministic. Although noise can be highly beneficial to the self-organization of complex systems~\citep{Kahneman2022}, our focus in this study was specifically on the relationship between inter-agent dynamics and consensus, rather than exploring how noise might modulate these dynamics.

We ran two groups of multi-agent experiments.
In the first group, to study the influence of the intra- and inter-agent coordination regimes, we varied the degree of internal coupling between the motor oscillators ($a_{v_i,v_j}$ values from 0 to 1, in steps of 0.02, in Eq.~\ref{eq:HKBsocial}), the social sensitivity ($S$ values from 0 to 5, in steps of 0.1, in Eq. \ref{eq:social_stimulus}), and the stimulus sensitivity ($c$ values from 0 and 10, in steps of 0.5, in Eq.~\ref{eq:HKBsocial}). In these experiments, the starting angle between agents was 10° and the brightness (i.e., quality) ratio of the two stimulus sources was $r = 0.8$. This ratio is lower than that in the single-agent case, to facilitate a wider range of collective behaviors. With a higher ratio, agents initially oriented towards the least bright stimulus source did not deviate enough from their initial movement path for collective dynamics to occur.

In the second group, to study the influence of the environmental and initial conditions, we varied the brightness (i.e., quality) ratio of the two stimulus sources ($r$ values from 0 to 1, in steps of 0.02) and the starting angles of the agents (from 0° to 18°, in steps of 0.36°). In these experiments, the stimulus sensitivity was $c=3$, social sensitivity is $S=1$, and internal connection was $a_{v_i,v_j}=0.5$.

\subsection{Evaluation}

In single-agent setups, performance is based on the agent's end position relative to a stimulus source. Note that, in these experiments, the agent could continue moving after reaching a source, so the performance metric includes how well the agent remained close to a source after initially approaching it.
For gradient ascent, we evaluated performance based on how closely the agent approaches the stimulus source:
\begin{equation}
   \texttt{performance} = 1 - \frac{D(t_{\texttt{end}})}{D(t_{0})}~,
\end{equation}
where $D(t_{\texttt{0}})$ and $D(t_{\texttt{end}})$ represent the agent's distance to the stimulus source at the beginning and end of the simulation.
For binary decision making, we evaluate how closely the agent approaches its closest stimulus source, regardless of the source's brightness level:
\begin{equation}
     \texttt{performance} = 1 - \frac{\min \left\{ D_{\texttt{source}_1}(t_{\texttt{end}}), D_{\texttt{source}_1}(t_{\texttt{end}})\right\} }{D(t_{0})}~,
\end{equation}
where $D_{\texttt{source}_1}(t_{\texttt{end}})$ and $D_{\texttt{source}_2}(t_{\texttt{end}})$ represent the agent's distance to $\texttt{source}_1$ and $\texttt{source}_2$ at the end of the simulation.
In multi-agent setups, performance is based on whether agents reach a consensus. Note that, 
in these experiments, an agent could no longer move once it came within 5 cm of a stimulus source, so any changes in agent angle and position due to agents circling around the stimulus source after arrival did not influence the decisions of the other agents. We evaluated performance based on the smallest average distance to one of the two stimulus sources:
\begin{equation}
     \texttt{performance} = \min \left\{ \frac{1}{N}\sum_{n=1}^{N}{\left[ 1 - \frac{D_{\texttt{source}_1}^n(t_{\texttt{end}})}{D(t_{0})} \right]}, \frac{1}{N}\sum_{n=1}^{N}{ \left[1 - \frac{ D_{\texttt{source}_2}^{n}(t_{\texttt{end}})}{D(t_{0})} \right]} \right\}~,
\end{equation}
with $N$ being the number of agents and $D_{\texttt{source}_1}^{n}(t)$ being the Euclidean distance from ${\texttt{source}_1}$ to agent $n$ at time $t$.

\subsubsection{Measures of coordination dynamics}
To evaluate the intra-agent neural dynamics, inter-agent neural dynamics, and inter-agent behavioral (i.e., movement) dynamics, we used the following measures: Kuramoto order parameters (KOP)~\citep{Strogatz2000}, phase locking value (PLV)~\citep{Lachaux1999}, and weighted phase-lag index (wPLI)~\citep{Vinck2011}.

First, we calculated KOP (i.e., the parameter $R(t)$~\citep{Strogatz2000}) as 
\begin{equation}
    Z(t) = R(t)e^{i \Theta t} = \frac{1}{N} \sum_{i=1}^{N}{e^{i \varphi_{i}(t)}}~,
\end{equation}
where $N$ is the number of oscillators and $\varphi_{i}$ is the phase angle of each oscillator (which, in this study, can be either the oscillator nodes of the neural controller or the movement directions of agents).
KOP quantifies the extent to which several oscillating components are in phase. If KOP is 1, all components are completely in phase, whereas low KOP values indicates an absence of synchronization between components. KOP values remaining constant over time indicate that the system has resorted to a stable dynamic (whether synchronized or not), whereas variation in the parameter indicate that the system is passing through various coordination states. Therefore, the standard deviation (SD) of KOP can be used as a measure of the metastability of coordination between oscillating components~\citep{Shanahan2010, Cabral2022}. We used metastability to assess the agents' neural dynamics but also to evaluate the collective movements in the multi-agent simulations. In the latter case, we used the measure of metastability to quantify the degree of `alignment variability', with which we mean the degree to which the collective switches between aligned and unaligned movement directions.

Based on~\cite{Lachaux1999}, we calculated sliding $\texttt{PLV}_{ij}$ for the connection between $v_i$ and $v_j$ as 
\begin{equation}
   \texttt{PLV}_{ij} =  \frac{1}{T} \left| \sum_{t=1}^{T}{e^{i \phi_{ij}(t)}} \right|~,
\end{equation}
where $T$ is the number of samples in a window. 
PLV is different from KOP in that it is maximal if the phases of the two oscillators are `locked', i.e., the relative phase of oscillators remains constant over time. We use PLV as a measure to indicate the degree of \textit{integration} of the oscillating components, i.e., the degree to which their phases are co-determined. PLV is often used as a measure of connectivity between brain components~\citep{Varela2001} as well as functional connectivity between the brains of different individuals during social interaction~\citep{Dumas2010}.


Finally, $\texttt{wPLI}_{ij}$ for the connection between $v_i$ and $v_j$ is calculated as follows~\citep{Vinck2011}:
\begin{equation}
   \texttt{wPLI}_{ij} = \frac{\frac{1}{T} \left| \sum_{t=1}^{T}{\left| I_{ij}(t) \right| \texttt{sgn}(I_{ij}(t))} \right|}{ \frac{1}{T}  \sum_{t=1}^{T}{\left| I_{ij}(t) \right|}}~,
\end{equation}
with
\begin{equation}
   I_{ij}(t) =   \texttt{Imag}({e^{i \phi_{ij}(t)}})~.
\end{equation}
Like the PLV, the wPLI characterizes to what degree different oscillators are integrated~\citep{Vinck2011} and has been used in several hyperscanning studies to quantify synchronization between brain regions~\citep[e.g.,][]{Schwartz2022}. 
wPLI differs from PLV in that its weighting of phases puts more emphasis on the covariance of phases than simple `locking'.
When zero-phase locking (i.e., completely synchronized activity) driven by common input needs to be distinguished from locking between other phases, wPLI provides a more robust characterization of oscillator integration than PLV.

\section*{Acknowledgments}
We thank Andreagiovanni Reina for useful suggestions to the manuscript. This work was partially supported by the program of Concerted Research Actions (ARC) of the Universit\'{e} libre de Bruxelles and a Mitacs Globalink Research Award to N. Coucke. M.K. Heinrich, A. Cleeremans, and M. Dorigo acknowledge support from the F.R.S.-FNRS, of which they are, respectively, postdoctoral researcher and research directors. 

\appendix

\bibliographystyle{unsrtnat}
\bibliography{references}  

\begin{thebibliography}{81}
\providecommand{\natexlab}[1]{#1}
\providecommand{\url}[1]{\texttt{#1}}
\expandafter\ifx\csname urlstyle\endcsname\relax
  \providecommand{\doi}[1]{doi: #1}\else
  \providecommand{\doi}{doi: \begingroup \urlstyle{rm}\Url}\fi

\bibitem[Bang and Frith(2017)]{Bang2017}
Dan Bang and Chris~D. Frith.
\newblock Making better decisions in groups.
\newblock \emph{Royal Society Open Science}, 4\penalty0 (8):\penalty0 170193,
  aug 2017.
\newblock \doi{10.1098/rsos.170193}.

\bibitem[Conradt and List(2008)]{Conradt2008}
Larissa Conradt and Christian List.
\newblock Group decisions in humans and animals: a survey.
\newblock \emph{Philosophical Transactions of the Royal Society B},
  364\penalty0 (1518):\penalty0 719--742, dec 2008.
\newblock \doi{10.1098/rstb.2008.0276}.

\bibitem[Hamann et~al.(2010)Hamann, Schmickl, Wörn, and
  Crailsheim]{Hamann2010}
Heiko Hamann, Thomas Schmickl, Heinz Wörn, and Karl Crailsheim.
\newblock Analysis of emergent symmetry breaking in collective decision making.
\newblock \emph{Neural Computing and Applications}, 21\penalty0 (2):\penalty0
  207--218, apr 2010.
\newblock \doi{10.1007/s00521-010-0368-6}.

\bibitem[{Montes de Oca} et~al.(2011){Montes de Oca}, Ferrante, Scheidler,
  Pinciroli, Birattari, and Dorigo]{MonFerSch-etal2011:si}
Marco~A. {Montes de Oca}, Eliseo Ferrante, Alexander Scheidler, Carlo
  Pinciroli, Mauro Birattari, and Marco Dorigo.
\newblock Majority-rule opinion dynamics with differential latency: A mechanism
  for self-organized collective decision-making.
\newblock \emph{Swarm Intelligence}, 5\penalty0 (3--4):\penalty0 305--327,
  2011.

\bibitem[Valentini et~al.(2017)Valentini, Ferrante, and Dorigo]{Valentini2017}
Gabriele Valentini, Eliseo Ferrante, and Marco Dorigo.
\newblock The best-of-n problem in robot swarms: Formalization, state of the
  art, and novel perspectives.
\newblock \emph{Frontiers in Robotics and {AI}}, 4, mar 2017.
\newblock \doi{10.3389/frobt.2017.00009}.

\bibitem[Couzin et~al.(2011)Couzin, Ioannou, Demirel, Gross, Torney, Hartnett,
  Conradt, Levin, and Leonard]{Couzin2011}
Iain~D. Couzin, Christos~C. Ioannou, Güven Demirel, Thilo Gross, Colin~J.
  Torney, Andrew Hartnett, Larissa Conradt, Simon~A. Levin, and Naomi~E.
  Leonard.
\newblock Uninformed individuals promote democratic consensus in animal groups.
\newblock \emph{Science}, 334\penalty0 (6062):\penalty0 1578--1580, dec 2011.
\newblock \doi{10.1126/science.1210280}.

\bibitem[Becker et~al.(2017)Becker, Brackbill, and Centola]{Becker2017}
Joshua Becker, Devon Brackbill, and Damon Centola.
\newblock Network dynamics of social influence in the wisdom of crowds.
\newblock \emph{Proceedings of the National Academy of Sciences}, page
  201615978, jun 2017.
\newblock \doi{10.1073/pnas.1615978114}.

\bibitem[Centola(2022)]{centola2022network}
Damon Centola.
\newblock The network science of collective intelligence.
\newblock \emph{Trends in Cognitive Sciences}, 26\penalty0 (11):\penalty0
  923--941, 2022.

\bibitem[Couzin et~al.(2005)Couzin, Krause, Franks, and Levin]{Couzin2005}
Iain~D. Couzin, Jens Krause, Nigel~R. Franks, and Simon~A. Levin.
\newblock Effective leadership and decision-making in animal groups on the
  move.
\newblock \emph{Nature}, 433\penalty0 (7025):\penalty0 513--516, feb 2005.
\newblock \doi{10.1038/nature03236}.

\bibitem[Yoo et~al.(2021)Yoo, Hayden, and Pearson]{Yoo2021}
Seng Bum~Michael Yoo, Benjamin~Yost Hayden, and John~M. Pearson.
\newblock Continuous decisions.
\newblock \emph{Philosophical Transactions of the Royal Society B: Biological
  Sciences}, 376\penalty0 (1819):\penalty0 20190664, jan 2021.
\newblock \doi{10.1098/rstb.2019.0664}.

\bibitem[Suzuki et~al.(2015)Suzuki, Adachi, Dunne, Bossaerts, and
  O'Doherty]{Suzuki2015}
Shinsuke Suzuki, Ryo Adachi, Simon Dunne, Peter Bossaerts, and John~P.
  O'Doherty.
\newblock Neural mechanisms underlying human consensus decision-making.
\newblock \emph{Neuron}, 86\penalty0 (2):\penalty0 591--602, apr 2015.
\newblock \doi{10.1016/j.neuron.2015.03.019}.

\bibitem[Hegselmann and Krause(2002)]{Hegselmann2002}
Rainer Hegselmann and Ulrich Krause.
\newblock Opinion dynamics and bounded confidence models, analysis and
  simulation.
\newblock \emph{Journal of Artificial Societies and Social Simulation}, 5, 07
  2002.

\bibitem[Galam(2008)]{Galam2008}
Serge Galam.
\newblock Sociophysics: A review of galam models.
\newblock \emph{International Journal of Modern Physics C}, March 2008.
\newblock \doi{10.1142/S0129183108012297}.

\bibitem[Flache et~al.(2017)Flache, Mäs, Feliciani, Chattoe-Brown, Deffuant,
  Huet, and Lorenz]{Flache2017}
Andreas Flache, Michael Mäs, Thomas Feliciani, Edmund Chattoe-Brown, Guillaume
  Deffuant, Sylvie Huet, and Jan Lorenz.
\newblock Models of social influence: Towards the next frontiers.
\newblock \emph{Journal of Artificial Societies and Social Simulation},
  20\penalty0 (4), 2017.
\newblock \doi{10.18564/jasss.3521}.

\bibitem[Srivastava and Leonard(2014)]{Srivastava2014}
Vaibhav Srivastava and Naomi~Ehrich Leonard.
\newblock Collective decision-making in ideal networks: The speed-accuracy
  tradeoff.
\newblock \emph{{IEEE} Transactions on Control of Network Systems}, 1\penalty0
  (1):\penalty0 121--132, mar 2014.
\newblock \doi{10.1109/tcns.2014.2310271}.

\bibitem[Tump et~al.(2020)Tump, Pleskac, and Kurvers]{Tump2020}
Alan~N. Tump, Timothy~J. Pleskac, and Ralf H. J.~M. Kurvers.
\newblock Wise or mad crowds? the cognitive mechanisms underlying information
  cascades.
\newblock \emph{Science Advances}, 6\penalty0 (29), jul 2020.
\newblock \doi{10.1126/sciadv.abb0266}.

\bibitem[Lokesh et~al.(2022)Lokesh, Sullivan, Calalo, Roth, Swanik, Carter, and
  Cashaback]{Lokesh2022}
Rakshith Lokesh, Seth Sullivan, Jan~A. Calalo, Adam Roth, Brenden Swanik,
  Michael~J. Carter, and Joshua G.~A. Cashaback.
\newblock Humans utilize sensory evidence of others' intended action to make
  online decisions.
\newblock \emph{Scientific Reports}, 12\penalty0 (1), may 2022.
\newblock \doi{10.1038/s41598-022-12662-y}.

\bibitem[Reina et~al.(2023)Reina, Bose, Srivastava, and Marshall]{Reina2023}
Andreagiovanni Reina, Thomas Bose, Vaibhav Srivastava, and James A.~R.
  Marshall.
\newblock Asynchrony rescues statistically optimal group decisions from
  information cascades through emergent leaders.
\newblock \emph{Royal Society Open Science}, 10\penalty0 (3), mar 2023.
\newblock \doi{10.1098/rsos.230175}.

\bibitem[Vicsek et~al.(1995)Vicsek, Czir{\'{o}}k, Ben-Jacob, Cohen, and
  Shochet]{Vicsek1995}
Tam{\'{a}}s Vicsek, Andr{\'{a}}s Czir{\'{o}}k, Eshel Ben-Jacob, Inon Cohen, and
  Ofer Shochet.
\newblock Novel type of phase transition in a system of self-driven particles.
\newblock \emph{Physical Review Letters}, 75\penalty0 (6):\penalty0 1226--1229,
  aug 1995.
\newblock \doi{10.1103/physrevlett.75.1226}.

\bibitem[Helbing and Molnar(1995)]{helbing1995social}
Dirk Helbing and Peter Molnar.
\newblock Social force model for pedestrian dynamics.
\newblock \emph{Physical review E}, 51\penalty0 (5):\penalty0 4282, 1995.

\bibitem[Ma et~al.(2016)Ma, Yuen, and Lee]{Ma2016}
Yi~Ma, Richard Kwok~Kit Yuen, and Eric Wai~Ming Lee.
\newblock Effective leadership for crowd evacuation.
\newblock \emph{Physica A: Statistical Mechanics and its Applications},
  450:\penalty0 333--341, may 2016.
\newblock \doi{10.1016/j.physa.2015.12.103}.

\bibitem[Reina et~al.(2017)Reina, Marshall, Trianni, and Bose]{reina2017model}
Andreagiovanni Reina, James~AR Marshall, Vito Trianni, and Thomas Bose.
\newblock Model of the best-of-n nest-site selection process in honeybees.
\newblock \emph{Physical Review E}, 95\penalty0 (5):\penalty0 052411, 2017.

\bibitem[Moussa{\"\i}d et~al.(2011)Moussa{\"\i}d, Helbing, and
  Theraulaz]{moussaid2011simple}
Mehdi Moussa{\"\i}d, Dirk Helbing, and Guy Theraulaz.
\newblock How simple rules determine pedestrian behavior and crowd disasters.
\newblock \emph{Proceedings of the National Academy of Sciences}, 108\penalty0
  (17):\penalty0 6884--6888, 2011.

\bibitem[Kingsbury and Hong(2020)]{kingsbury2020multi}
Lyle Kingsbury and Weizhe Hong.
\newblock A multi-brain framework for social interaction.
\newblock \emph{Trends in neurosciences}, 43\penalty0 (9):\penalty0 651--666,
  2020.

\bibitem[Moreau and Dumas(2021)]{moreau2021beyond}
Quentin Moreau and Guillaume Dumas.
\newblock Beyond “correlation vs. causation”: multi-brain neuroscience
  needs explanation.
\newblock \emph{Trends Cogn. Sci}, 25:\penalty0 542--543, 2021.

\bibitem[Res{\'e}ndiz-Benhumea et~al.(2021)Res{\'e}ndiz-Benhumea, Sangati,
  Sangati, Keshmiri, and Froese]{resendiz2021shrunken}
Georgina~Montserrat Res{\'e}ndiz-Benhumea, Ekaterina Sangati, Federico Sangati,
  Soheil Keshmiri, and Tom Froese.
\newblock Shrunken social brains? a minimal model of the role of social
  interaction in neural complexity.
\newblock \emph{Frontiers in Neurorobotics}, 15:\penalty0 634085, 2021.

\bibitem[Sridhar et~al.(2021)Sridhar, Li, Gorbonos, Nagy, Schell, Sorochkin,
  Gov, and Couzin]{Sridhar2021}
Vivek~H. Sridhar, Liang Li, Dan Gorbonos, M{\'{a}}t{\'{e}} Nagy, Bianca~R.
  Schell, Timothy Sorochkin, Nir~S. Gov, and Iain~D. Couzin.
\newblock The geometry of decision-making in individuals and collectives.
\newblock \emph{Proceedings of the National Academy of Sciences}, 118\penalty0
  (50), dec 2021.
\newblock \doi{10.1073/pnas.2102157118}.

\bibitem[Heins et~al.(2024)Heins, Millidge, Da~Costa, Mann, Friston, and
  Couzin]{heins2024collective}
Conor Heins, Beren Millidge, Lancelot Da~Costa, Richard~P Mann, Karl~J Friston,
  and Iain~D Couzin.
\newblock Collective behavior from surprise minimization.
\newblock \emph{Proceedings of the National Academy of Sciences}, 121\penalty0
  (17):\penalty0 e2320239121, 2024.

\bibitem[Clark(1998)]{Clark1998}
Andy Clark.
\newblock \emph{Being There}.
\newblock The MIT Press, 1998.
\newblock ISBN 9780262531566.

\bibitem[Newen and Gallagher(2018)]{Newen2018}
Albert Newen and Shaun Gallagher.
\newblock \emph{The Oxford Handbook of 4E Cognition}.
\newblock Oxford University Press, 2018.
\newblock ISBN 9780198735410.

\bibitem[Steels and Brooks(2018)]{Steels2018}
Luc Steels and Rodney Brooks.
\newblock \emph{The artificial life route to artificial intelligence: Building
  embodied, situated agents}.
\newblock Routledge, 2018.

\bibitem[Colas et~al.(2022)Colas, Karch, Moulin-Frier, and Oudeyer]{Colas2022}
C{\'{e}}dric Colas, Tristan Karch, Cl{\'{e}}ment Moulin-Frier, and Pierre-Yves
  Oudeyer.
\newblock Language and culture internalization for human-like autotelic {AI}.
\newblock \emph{Nature Machine Intelligence}, 4\penalty0 (12):\penalty0
  1068--1076, dec 2022.
\newblock \doi{10.1038/s42256-022-00591-4}.

\bibitem[Varela et~al.(1992)Varela, Thompson, and Rosch]{Varela1992}
Francisco~J. Varela, Evan~T. Thompson, and Eleanor Rosch.
\newblock \emph{The Embodied Mind}.
\newblock The MIT Press, 1992.
\newblock ISBN 9780262720212.

\bibitem[Thompson and Varela(2001)]{Thompson2001}
Evan Thompson and Francisco~J. Varela.
\newblock Radical embodiment: neural dynamics and consciousness.
\newblock \emph{Trends in Cognitive Sciences}, 5\penalty0 (10):\penalty0
  418--425, oct 2001.
\newblock \doi{10.1016/s1364-6613(00)01750-2}.

\bibitem[Froese and Paolo(2011)]{Froese2011}
Tom Froese and Ezequiel A.~Di Paolo.
\newblock The enactive approach.
\newblock \emph{Pragmatics and Cognition}, 19\penalty0 (1):\penalty0 1--36, jul
  2011.
\newblock \doi{10.1075/pc.19.1.01fro}.

\bibitem[Ward(2003)]{ward2003synchronous}
Lawrence~M Ward.
\newblock Synchronous neural oscillations and cognitive processes.
\newblock \emph{Trends in cognitive sciences}, 7\penalty0 (12):\penalty0
  553--559, 2003.

\bibitem[Buzs{\'{a}}ki(2019)]{Buzsaki2019}
György Buzs{\'{a}}ki.
\newblock \emph{The Brain from Inside Out}.
\newblock Oxford University Press, jun 2019.
\newblock \doi{10.1093/oso/9780190905385.001.0001}.

\bibitem[Buzs{\'{a}}ki et~al.(2013)Buzs{\'{a}}ki, Logothetis, and
  Singer]{Buzsaki2013}
György Buzs{\'{a}}ki, Nikos Logothetis, and Wolf Singer.
\newblock Scaling brain size, keeping timing: Evolutionary preservation of
  brain rhythms.
\newblock \emph{Neuron}, 80\penalty0 (3):\penalty0 751--764, oct 2013.
\newblock \doi{10.1016/j.neuron.2013.10.002}.

\bibitem[Senoussi et~al.(2022)Senoussi, Verbeke, Desender, De~Loof, Talsma, and
  Verguts]{senoussi2022theta}
Mehdi Senoussi, Pieter Verbeke, Kobe Desender, Esther De~Loof, Durk Talsma, and
  Tom Verguts.
\newblock Theta oscillations shift towards optimal frequency for cognitive
  control.
\newblock \emph{Nature Human Behaviour}, 6\penalty0 (7):\penalty0 1000--1013,
  2022.

\bibitem[Charalambous and Djebbara(2023)]{charalambous2023natural}
Efrosini Charalambous and Zakaria Djebbara.
\newblock On natural attunement: shared rhythms between the brain and the
  environment.
\newblock \emph{Neuroscience \& Biobehavioral Reviews}, 155:\penalty0 105438,
  2023.

\bibitem[Kelso(1997)]{Kelso1997}
J.~A.~Scott Kelso.
\newblock \emph{Dynamic Patterns The Self-organization Of Brain And Behavior}.
\newblock Bradford Book, 1997.
\newblock ISBN 9780262611312.

\bibitem[Kelso et~al.(2014)Kelso, Tognoli, and Dumas]{Kelso2014}
J.~A.~Scott Kelso, Emmanuelle Tognoli, and Guillaume Dumas.
\newblock Coordination dynamics: Bidirectional coupling between humans,
  machines and brains.
\newblock In \emph{2014 {IEEE} International Conference on Systems, Man, and
  Cybernetics ({SMC})}. {IEEE}, oct 2014.
\newblock \doi{10.1109/smc.2014.6974258}.

\bibitem[Tognoli et~al.(2020)Tognoli, Zhang, Fuchs, Beetle, and
  Kelso]{Tognoli2020}
Emmanuelle Tognoli, Mengsen Zhang, Armin Fuchs, Christopher Beetle, and
  J.~A.~Scott Kelso.
\newblock Coordination dynamics: A foundation for understanding social
  behavior.
\newblock \emph{Frontiers in Human Neuroscience}, 14, aug 2020.
\newblock \doi{10.3389/fnhum.2020.00317}.

\bibitem[Tognoli and Kelso(2014)]{Tognoli2014}
Emmanuelle Tognoli and J.~A.~Scott Kelso.
\newblock The metastable brain.
\newblock \emph{Neuron}, 81\penalty0 (1):\penalty0 35--48, jan 2014.
\newblock \doi{10.1016/j.neuron.2013.12.022}.

\bibitem[Haken et~al.(1985)Haken, Kelso, and Bunz]{Haken1985}
Hermann Haken, J.~A.~Scott Kelso, and H.~Bunz.
\newblock A theoretical model of phase transitions in human hand movements.
\newblock \emph{Biological Cybernetics}, 51\penalty0 (5):\penalty0 347--356,
  feb 1985.
\newblock \doi{10.1007/bf00336922}.

\bibitem[Kuramoto(1984)]{Kuramoto1984}
Y.~Kuramoto.
\newblock \emph{Chemical Oscillations, Waves, and Turbulence}.
\newblock Springer Berlin Heidelberg, 1984.
\newblock ISBN 9783642696916.

\bibitem[Kelso(2013)]{Kelso2013}
J.~A.~Scott Kelso.
\newblock Coordination dynamics.
\newblock In \emph{Encyclopedia of Complexity and Systems Science}, pages
  1--41. Springer New York, 2013.
\newblock \doi{10.1007/978-3-642-27737-5_101-3}.

\bibitem[Aguilera et~al.(2013)Aguilera, Bedia, Santos, and
  Barandiaran]{Aguilera2013}
Miguel Aguilera, Manuel~G. Bedia, Bruno~A. Santos, and Xabier~E. Barandiaran.
\newblock The situated {HKB} model: how sensorimotor spatial coupling can alter
  oscillatory brain dynamics.
\newblock \emph{Frontiers in Computational Neuroscience}, 7, 2013.
\newblock \doi{10.3389/fncom.2013.00117}.

\bibitem[Leonard et~al.(2011)Leonard, Shen, Nabet, Scardovi, Couzin, and
  Levin]{Leonard2011}
Naomi~E. Leonard, Tian Shen, Benjamin Nabet, Luca Scardovi, Iain~D. Couzin, and
  Simon~A. Levin.
\newblock Decision versus compromise for animal groups in motion.
\newblock \emph{Proceedings of the National Academy of Sciences}, 109\penalty0
  (1):\penalty0 227--232, dec 2011.
\newblock \doi{10.1073/pnas.1118318108}.

\bibitem[Braitenberg(1986)]{Braitenberg1986}
Valentino Braitenberg.
\newblock \emph{Vehicles, Experiments in Synthetic Psychology.}
\newblock M.I.T. P., 1986.
\newblock ISBN 0262521121.

\bibitem[Zhang et~al.(2019)Zhang, Beetle, Kelso, and Tognoli]{Zhang2019}
Mengsen Zhang, Christopher Beetle, J.~A.~Scott Kelso, and Emmanuelle Tognoli.
\newblock Connecting empirical phenomena and theoretical models of biological
  coordination across scales.
\newblock \emph{Journal of The Royal Society Interface}, 16\penalty0
  (157):\penalty0 20190360, aug 2019.
\newblock \doi{10.1098/rsif.2019.0360}.

\bibitem[Varela et~al.(2001)Varela, Lachaux, Rodriguez, and
  Martinerie]{Varela2001}
Francisco Varela, Jean-Philippe Lachaux, Eugenio Rodriguez, and Jacques
  Martinerie.
\newblock The brainweb: Phase synchronization and large-scale integration.
\newblock \emph{Nature Reviews Neuroscience}, 2\penalty0 (4):\penalty0
  229--239, apr 2001.
\newblock \doi{10.1038/35067550}.

\bibitem[Avena-Koenigsberger et~al.(2018)Avena-Koenigsberger, Misic, and
  Sporns]{avena2018communication}
Andrea Avena-Koenigsberger, Bratislav Misic, and Olaf Sporns.
\newblock Communication dynamics in complex brain networks.
\newblock \emph{Nature reviews neuroscience}, 19\penalty0 (1):\penalty0 17--33,
  2018.

\bibitem[Strogatz(2000)]{Strogatz2000}
Steven~H. Strogatz.
\newblock From kuramoto to crawford: exploring the onset of synchronization in
  populations of coupled oscillators.
\newblock \emph{Physica D: Nonlinear Phenomena}, 143\penalty0 (1-4):\penalty0
  1--20, sep 2000.
\newblock \doi{10.1016/s0167-2789(00)00094-4}.

\bibitem[Cabral et~al.(2022)Cabral, Castaldo, Vohryzek, Litvak, Bick,
  Lambiotte, Friston, Kringelbach, and Deco]{Cabral2022}
Joana Cabral, Francesca Castaldo, Jakub Vohryzek, Vladimir Litvak, Christian
  Bick, Renaud Lambiotte, Karl Friston, Morten~L. Kringelbach, and Gustavo
  Deco.
\newblock Metastable oscillatory modes emerge from synchronization in the brain
  spacetime connectome.
\newblock \emph{Communications Physics}, 5\penalty0 (1), jul 2022.
\newblock \doi{10.1038/s42005-022-00950-y}.

\bibitem[Vinck et~al.(2011)Vinck, Oostenveld, van Wingerden, Battaglia, and
  Pennartz]{Vinck2011}
Martin Vinck, Robert Oostenveld, Marijn van Wingerden, Franscesco Battaglia,
  and Cyriel~M.A. Pennartz.
\newblock An improved index of phase-synchronization for electrophysiological
  data in the presence of volume-conduction, noise and sample-size bias.
\newblock \emph{{NeuroImage}}, 55\penalty0 (4):\penalty0 1548--1565, apr 2011.
\newblock \doi{10.1016/j.neuroimage.2011.01.055}.

\bibitem[Czeszumski et~al.(2020)Czeszumski, Eustergerling, Lang, Menrath,
  Gerstenberger, Schuberth, Schreiber, Rendon, and König]{Czeszumski2020}
Artur Czeszumski, Sara Eustergerling, Anne Lang, David Menrath, Michael
  Gerstenberger, Susanne Schuberth, Felix Schreiber, Zadkiel~Zuluaga Rendon,
  and Peter König.
\newblock Hyperscanning: A valid method to study neural inter-brain
  underpinnings of social interaction.
\newblock \emph{Frontiers in Human Neuroscience}, 14, feb 2020.
\newblock \doi{10.3389/fnhum.2020.00039}.

\bibitem[Schwartz et~al.(2022)Schwartz, Levy, Endevelt-Shapira, Djalovski,
  Hayut, Dumas, and Feldman]{Schwartz2022}
Linoy Schwartz, Jonathan Levy, Yaara Endevelt-Shapira, Amir Djalovski, Olga
  Hayut, Guillaume Dumas, and Ruth Feldman.
\newblock Technologically-assisted communication attenuates inter-brain
  synchrony.
\newblock \emph{{NeuroImage}}, 264:\penalty0 119677, dec 2022.
\newblock \doi{10.1016/j.neuroimage.2022.119677}.

\bibitem[Waller(2008)]{Waller2008}
John Waller.
\newblock \emph{A time to dance, a time to die}.
\newblock Icon Books, 2008.
\newblock ISBN 9781848310216.

\bibitem[Couzin and Franks(2003)]{Couzin2003a}
I.~D. Couzin and N.~R. Franks.
\newblock Self-organized lane formation and optimized traffic flow in army
  ants.
\newblock \emph{Proceedings of the Royal Society of London. Series B:
  Biological Sciences}, 270\penalty0 (1511):\penalty0 139--146, jan 2003.
\newblock \doi{10.1098/rspb.2002.2210}.

\bibitem[Valentini et~al.(2015)Valentini, Ferrante, Hamann, and
  Dorigo]{Valentini2015}
Gabriele Valentini, Eliseo Ferrante, Heiko Hamann, and Marco Dorigo.
\newblock Collective decision with 100~kilobots: speed versus accuracy in
  binary discrimination problems.
\newblock \emph{Autonomous Agents and Multi-Agent Systems}, 30\penalty0
  (3):\penalty0 553--580, dec 2015.
\newblock \doi{10.1007/s10458-015-9323-3}.

\bibitem[Hölldobler and Wilson(1990)]{Hoelldobler1990}
Bert Hölldobler and Edward~Osborne Wilson.
\newblock \emph{The Ants}.
\newblock Belknap Press, 1990.
\newblock ISBN 9780674040755.

\bibitem[Pickavance et~al.(2018)Pickavance, Azmoodeh, and
  Wilson]{Pickavance2018}
John Pickavance, Arianne Azmoodeh, and Andrew~D. Wilson.
\newblock The effects of feedback format, and egocentric \& allocentric
  relative phase on coordination stability.
\newblock \emph{Human Movement Science}, 59:\penalty0 143--152, jun 2018.
\newblock \doi{10.1016/j.humov.2018.04.005}.

\bibitem[Eric~Bonabeau(1999)]{Bonabeau1999}
Guy~Theraulaz Eric~Bonabeau, Marco~Dorigo.
\newblock \emph{Swarm Intelligence}.
\newblock Oxford University Press Inc, 1999.
\newblock ISBN 0195131592.
\newblock URL
  \url{https://www.ebook.de/de/product/3253990/eric_postdoctoral_fellow_postdoctoral_fellow_santa_fe_institute_bonabeau_marco_researcher_researcher_free_university_of_brussels_dorigo_guy_researcher_researcher_cnrs_university_paul_sabatier_theraulaz_swarm_intelligence.html}.

\bibitem[Kahneman et~al.(2022)Kahneman, Krakauer, Sibony, Sunstein, and
  Wolpert]{Kahneman2022}
Daniel Kahneman, David~C Krakauer, Olivier Sibony, Cass Sunstein, and David
  Wolpert.
\newblock An exchange of letters on the role of noise in collective
  intelligence.
\newblock \emph{Collective Intelligence}, 1\penalty0 (1):\penalty0
  263391372210785, aug 2022.
\newblock \doi{10.1177/26339137221078593}.

\bibitem[O'Keeffe et~al.(2017)O'Keeffe, Hong, and Strogatz]{OKeeffe2017}
Kevin~P. O'Keeffe, Hyunsuk Hong, and Steven~H. Strogatz.
\newblock Oscillators that sync and swarm.
\newblock \emph{Nature Communications}, 8\penalty0 (1), nov 2017.
\newblock \doi{10.1038/s41467-017-01190-3}.

\bibitem[Ceron et~al.(2023)Ceron, O'Keeffe, and Petersen]{Ceron2023}
Steven Ceron, Kevin O'Keeffe, and Kirstin Petersen.
\newblock Diverse behaviors in non-uniform chiral and non-chiral swarmalators.
\newblock \emph{Nature Communications}, 14\penalty0 (1), feb 2023.
\newblock \doi{10.1038/s41467-023-36563-4}.

\bibitem[Auvray et~al.(2009)Auvray, Lenay, and Stewart]{auvray2009perceptual}
Malika Auvray, Charles Lenay, and John Stewart.
\newblock Perceptual interactions in a minimalist virtual environment.
\newblock \emph{New ideas in psychology}, 27\penalty0 (1):\penalty0 32--47,
  2009.

\bibitem[{De Jaegher} et~al.(2010){De Jaegher}, {Di Paolo}, and
  Gallagher]{DeJaegher2010}
Hanne {De Jaegher}, Ezequiel {Di Paolo}, and Shaun Gallagher.
\newblock Can social interaction constitute social cognition?
\newblock \emph{Trends in Cognitive Sciences}, 14\penalty0 (10):\penalty0
  441--447, oct 2010.
\newblock \doi{10.1016/j.tics.2010.06.009}.

\bibitem[Dumas et~al.(2010)Dumas, Nadel, Soussignan, Martinerie, and
  Garnero]{Dumas2010}
Guillaume Dumas, Jacqueline Nadel, Robert Soussignan, Jacques Martinerie, and
  Line Garnero.
\newblock Inter-brain synchronization during social interaction.
\newblock \emph{{PLoS} {ONE}}, 5\penalty0 (8):\penalty0 e12166, aug 2010.
\newblock \doi{10.1371/journal.pone.0012166}.

\bibitem[Yang et~al.(2021)Yang, Wu, V{\'{a}}zquez-Guardado, Wegener,
  Grajales-Reyes, Deng, Wang, Avila, Moreno, Minkowicz, Dumrongprechachan, Lee,
  Zhang, Legaria, Ma, Mehta, Franklin, Hartman, Bai, Han, Zhao, Lu, Yu, Sheng,
  Banks, Yu, Donaldson, Gereau, Good, Xie, Huang, Kozorovitskiy, and
  Rogers]{Yang2021}
Yiyuan Yang, Mingzheng Wu, Abraham V{\'{a}}zquez-Guardado, Amy~J. Wegener,
  Jose~G. Grajales-Reyes, Yujun Deng, Taoyi Wang, Raudel Avila, Justin~A.
  Moreno, Samuel Minkowicz, Vasin Dumrongprechachan, Jungyup Lee, Shuangyang
  Zhang, Alex~A. Legaria, Yuhang Ma, Sunita Mehta, Daniel Franklin, Layne
  Hartman, Wubin Bai, Mengdi Han, Hangbo Zhao, Wei Lu, Yongjoon Yu, Xing Sheng,
  Anthony Banks, Xinge Yu, Zoe~R. Donaldson, Robert~W. Gereau, Cameron~H. Good,
  Zhaoqian Xie, Yonggang Huang, Yevgenia Kozorovitskiy, and John~A. Rogers.
\newblock Wireless multilateral devices for optogenetic studies of individual
  and social behaviors.
\newblock \emph{Nature Neuroscience}, 24\penalty0 (7):\penalty0 1035--1045, may
  2021.
\newblock \doi{10.1038/s41593-021-00849-x}.

\bibitem[Dumas et~al.(2012)Dumas, Chavez, Nadel, and Martinerie]{Dumas2012a}
Guillaume Dumas, Mario Chavez, Jacqueline Nadel, and Jacques Martinerie.
\newblock Anatomical connectivity influences both intra- and inter-brain
  synchronizations.
\newblock \emph{{PLoS} {ONE}}, 7\penalty0 (5):\penalty0 e36414, may 2012.
\newblock \doi{10.1371/journal.pone.0036414}.

\bibitem[Moreau et~al.(2022)Moreau, Adel, Douglas, Ranjbaran, and
  Dumas]{Moreau2022}
Quentin Moreau, Lena Adel, Caitriona Douglas, Ghazaleh Ranjbaran, and Guillaume
  Dumas.
\newblock A neurodynamic model of inter-brain coupling in the gamma band.
\newblock \emph{Journal of Neurophysiology}, sep 2022.
\newblock \doi{10.1152/jn.00224.2022}.

\bibitem[Heggli et~al.(2019)Heggli, Cabral, Konvalinka, Vuust, and
  Kringelbach]{Heggli2019}
Ole~Adrian Heggli, Joana Cabral, Ivana Konvalinka, Peter Vuust, and Morten~L.
  Kringelbach.
\newblock A kuramoto model of self-other integration across interpersonal
  synchronization strategies.
\newblock \emph{{PLOS} Computational Biology}, 15\penalty0 (10):\penalty0
  e1007422, oct 2019.
\newblock \doi{10.1371/journal.pcbi.1007422}.

\bibitem[Bolotta and Dumas(2022)]{Bolotta2021}
Samuele Bolotta and Guillaume Dumas.
\newblock Social {Neuro} {AI}: {Social} {Interaction} as the “{Dark}
  {Matter}” of {AI}.
\newblock \emph{Frontiers in Computer Science}, 4, 2022.
\newblock ISSN 2624-9898.
\newblock URL
  \url{https://www.frontiersin.org/articles/10.3389/fcomp.2022.846440}.

\bibitem[Van~Rossum and Drake(2009)]{Rossum}
Guido Van~Rossum and Fred~L. Drake.
\newblock \emph{Python 3 Reference Manual}.
\newblock CreateSpace, Scotts Valley, CA, 2009.
\newblock ISBN 1441412697.

\bibitem[Paszke et~al.(2019)Paszke, Gross, Massa, Lerer, Bradbury, Chanan,
  Killeen, Lin, Gimelshein, Antiga, Desmaison, Kopf, Yang, DeVito, Raison,
  Tejani, Chilamkurthy, Steiner, Fang, Bai, and Chintala]{PyTorch2019}
Adam Paszke, Sam Gross, Francisco Massa, Adam Lerer, James Bradbury, Gregory
  Chanan, Trevor Killeen, Zeming Lin, Natalia Gimelshein, Luca Antiga, Alban
  Desmaison, Andreas Kopf, Edward Yang, Zachary DeVito, Martin Raison, Alykhan
  Tejani, Sasank Chilamkurthy, Benoit Steiner, Lu~Fang, Junjie Bai, and Soumith
  Chintala.
\newblock Pytorch: An imperative style, high-performance deep learning library.
\newblock In \emph{Advances in Neural Information Processing Systems 32}, pages
  8024--8035. Curran Associates, Inc., 2019.
\newblock URL
  \url{http://papers.neurips.cc/paper/9015-pytorch-an-imperative-style-high-performance-deep-learning-library.pdf}.

\bibitem[Sterling and Laughlin(2017)]{Sterling2017}
Peter Sterling and Simon Laughlin.
\newblock \emph{Principles of Neural Design}.
\newblock The MIT Press, 2017.
\newblock ISBN 9780262534680.

\bibitem[Nabet et~al.(2009)Nabet, Leonard, Couzin, and Levin]{Nabet2009}
Benjamin Nabet, Naomi~E. Leonard, Iain~D. Couzin, and Simon~A. Levin.
\newblock Dynamics of decision making in animal group motion.
\newblock \emph{Journal of Nonlinear Science}, 19\penalty0 (4):\penalty0
  399--435, jan 2009.
\newblock ISSN 1432-1467.
\newblock \doi{10.1007/s00332-008-9038-6}.

\bibitem[Lachaux et~al.(1999)Lachaux, Rodriguez, Martinerie, and
  Varela]{Lachaux1999}
Jean-Philippe Lachaux, Eugenio Rodriguez, Jacques Martinerie, and Francisco~J.
  Varela.
\newblock Measuring phase synchrony in brain signals.
\newblock \emph{Human Brain Mapping}, 8\penalty0 (4):\penalty0 194--208, 1999.
\newblock \doi{10.1002/(sici)1097-0193(1999)8:4<194::aid-hbm4>3.0.co;2-c}.

\bibitem[Shanahan(2010)]{Shanahan2010}
Murray Shanahan.
\newblock Metastable chimera states in community-structured oscillator
  networks.
\newblock \emph{Chaos: An Interdisciplinary Journal of Nonlinear Science},
  20\penalty0 (1):\penalty0 013108, mar 2010.
\newblock \doi{10.1063/1.3305451}.

\end{thebibliography}


\begin{thebibliography}{1}

\bibitem{Aguilera2013}
Miguel Aguilera, Manuel~G. Bedia, Bruno~A. Santos, and Xabier~E. Barandiaran.
\newblock The situated {HKB} model: how sensorimotor spatial coupling can alter
  oscillatory brain dynamics.
\newblock {\em Frontiers in Computational Neuroscience}, 7, 2013.

\bibitem{Haken1985}
Hermann Haken, J.~A.~Scott Kelso, and H.~Bunz.
\newblock A theoretical model of phase transitions in human hand movements.
\newblock {\em Biological Cybernetics}, 51(5):347--356, feb 1985.

\bibitem{Kuramoto1984}
Y.~Kuramoto.
\newblock {\em Chemical Oscillations, Waves, and Turbulence}.
\newblock Springer Berlin Heidelberg, 1984.

\bibitem{McKinley2021}
Joseph McKinley, Mengsen Zhang, Alice Wead, Christine Williams, Emmanuelle
  Tognoli, and Christopher Beetle.
\newblock Third party stabilization of unstable coordination in systems of
  coupled oscillators.
\newblock {\em Journal of Physics: Conference Series}, 2090(1):012167, nov
  2021.

\bibitem{Schoener1988}
G.~Schöner and J.A.S. Kelso.
\newblock A dynamic pattern theory of behavioral change.
\newblock {\em Journal of Theoretical Biology}, 135(4):501--524, dec 1988.

\bibitem{Tognoli2014}
Emmanuelle Tognoli and J.~A.~Scott Kelso.
\newblock The metastable brain.
\newblock {\em Neuron}, 81(1):35--48, jan 2014.

\bibitem{Zhang2019}
Mengsen Zhang, Christopher Beetle, J.~A.~Scott Kelso, and Emmanuelle Tognoli.
\newblock Connecting empirical phenomena and theoretical models of biological
  coordination across scales.
\newblock {\em Journal of The Royal Society Interface}, 16(157):20190360, aug
  2019.

\end{thebibliography}

\end{document}




\maketitle


\section{The agent design and HKB equations}

Assume that an oscillator has intrinsic frequency $\omega$ and phase $\varphi$. In a system with two uncoupled oscillators, the phase angle between the oscillators is governed by 
\begin{equation}
    \dot{\phi} = \delta \omega~,
\end{equation}
where \(\delta \omega\) is the difference between the intrinsic frequency of the oscillators and \(\phi = \varphi_1 - \varphi_2\) is the instantaneous phase difference between the two oscillators.
Realistically, biological brain regions are coupled and influence each other's activity. 
The simplest possible interaction between two components is in-phase coupling, i.e., the oscillators will be attracted to synchronous oscillations. This relationship has been modeled by the Kuramoto equation~\cite{Kuramoto1984}:
\begin{equation}
    \dot{\phi} = \delta \omega - a \sin{\phi}~.
\end{equation}
Whether a stable coordination state exists for the system depends on the relative sizes of $\delta \omega$ and $a$; no stable in-phase coordination will occur when $\delta \omega$ is larger than $a$.

Biological components can also have a more complex form of coordination. To explain bistable coordination between hand movements and accommodates antiphase coupling between the oscillators, the Haken-Kelso-Bunz (HKB) equation was introduced~\cite{Haken1985}:
\begin{equation}
    \dot{\phi} = \delta \omega - a \sin{\phi} - 2 b \sin{2\phi}~,
\label{eq:HKB}
\end{equation}
where $a$ and $b$ determine the strength of phase and antiphase coupling, respectively. Whether stable antiphase coupling is present in this system depends on the ratio \(k=\frac{a}{b}\). Stable antiphase coupling is only possible when $k<4$. Note that even when no stable coordination pattern is present, both the phase and the antiphase coupling might still contribute to creating rich metastable dynamics in which the system continuously moves from phase to antiphase, with durations proportional to the coupling strengths~\cite{Tognoli2014}.

\subsection{Situated HKB agents}
The broader context in which a system is situated can profoundly affect the coordination behaviors it can exhibit~\cite{McKinley2021}. For example, adding a source of external influence can enhance and inhibit stable coordination regimes~\cite{Schoener1988}. 
The HKB equation can be applied to situated contexts and used to model a cognitive system's reaction to sensory inputs from the external environment. In the \textit{situated HKB agent} introduced by~\cite{Aguilera2013}, the HKB equation was used to model the interaction between sensory and motor brain regions of the agent. The agent was coupled to the environment so that its intra-agent neural dynamics (phase angle between sensory and motor regions) were modulated by sensory information, such that:
\begin{equation}
    \dot{\phi} = \delta \omega + c I - a \sin{\phi} - 2 b \sin{2\phi}~,
\end{equation}
where $I$ is the intensity of the stimulus and $c$ is the stimulus sensitivity of the agent.
A closed sensorimotor loop was created by controlling the agent's movement in the environment according to the phase angle between the two oscillators, such that
\begin{equation}
    \dot{\theta} = \eta \phi~,
\end{equation}
where $\theta$ is the orientation of the agent in the environment and $\eta$ is a scaling factor.

\subsubsection{HKB equations for \textit{N} components}
To enable modeling of $N+1$ interacting components, \cite{Zhang2019} introduced variations on the HKB equations in which the phase of each oscillator is explicitly modeled, rather than only the phase difference. After rescaling the parameters $a_{ij}$ and $b_{ij}$ (i.e., the parameters associated with the connections between $v_i$ and $v_j$), the HKB equation becomes:
\begin{equation}
    \dot{\varphi}_i = \delta \omega - \sum^{N}_{j=0} a_{ij} \sin{\phi_{ij}} - 2 b_{ij}  \sin{2\phi_{ij}}~,
\end{equation}
where $\omega_i$ is the intrinsic frequency of the $i^{th}$ component.

\subsection{Our controller design: social HKB agents}
Combining the equations above, we achieve the set of update equations reported in the main text:

\begin{equation} 
\dot{\boldsymbol\varphi} 
\begin{cases}
    {\dot{\varphi}_{v_1}} = {\omega_{v_1} + c I_l - \sum_{j} a_{v_1,v_j} \sin{\phi_{v_1,v_j}} - \sum_{j} b_{v_1,v_j} \sin{2\phi_{v_1,v_j}}},\\
    {\dot{\varphi}_{v_2}} = {\omega_{v_2} + c I_r -\sum_{j} a_{v_2,v_j} \sin{\phi_{v_2,v_j}} - \sum_{j} b_{v_2,v_j} \sin{2\phi_{v_2,v_j}}},\\
    {\dot{\varphi}_{v_3}} = \omega_{v_3} - \sum_{j} a_{v_3,v_j} \sin{\phi_{v_3,v_j}} - \sum_{j} b_{v_3,v_j} \sin{2\phi_{v_3,v_j}},\\
    {\dot{\varphi}_{v_4}} = \omega_{v_4} - \sum_{j} a_{v_4,v_j} \sin{\phi_{v_4,v_j}} - \sum_{j} b_{v_4,v_j} \sin{2\phi_{v_4,v_j}}\\
\end{cases},
\label{eq:HKBsocial}
\end{equation}

\subsection{Performance evaluation}

The performance for multi-agent consensus achievement is defined as:

\begin{equation*}
     \texttt{performance} = \min \left\{ \frac{1}{N}\sum_{n=1}^{N}{\left[ 1 - \frac{D_{\texttt{source}_1}^n(t_{\texttt{end}})}{D(t_{0})} \right]}, \frac{1}{N}\sum_{n=1}^{N}{ \left[1 - \frac{ D_{\texttt{source}_2}^{n}(t_{\texttt{end}})}{D(t_{0})} \right]} \right\}~,
\end{equation*}

with $N$ being the number of agents and $D_{\texttt{source}_1}^{n}(t)$ being the Euclidean distance from ${\texttt{source}_1}$ to agent $n$ at time $t$.

\begin{figure}
\centering
\includegraphics[width=\linewidth]{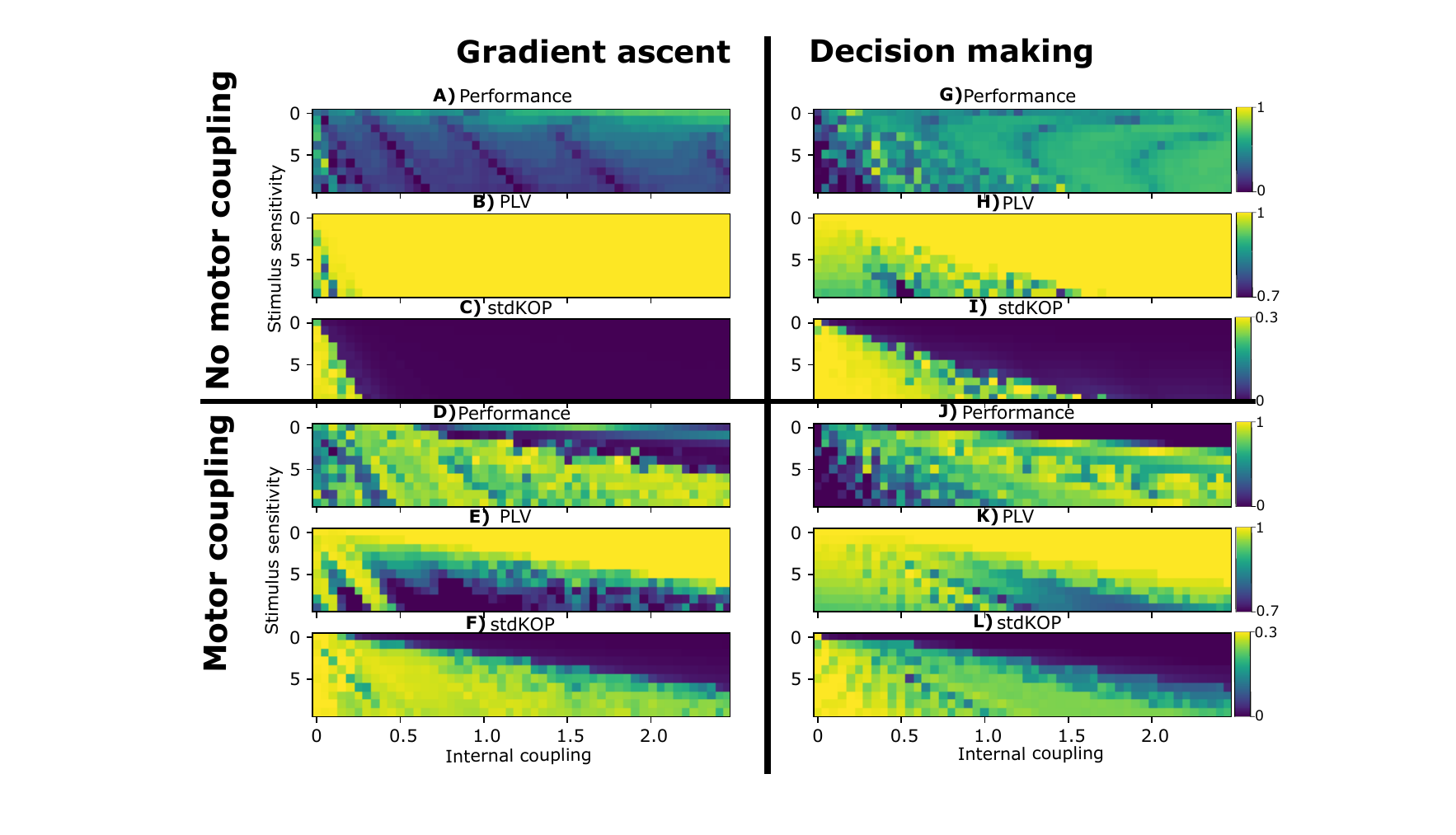} 
\caption{Agent performance and intra-agent neural dynamics in both types of single-agent experiments: (A-F) gradient ascent and (G-L) binary decision-making. Agent variations are either (D-F;J-L) with with motor coupling or (A-C;G-I) without motor coupling. In all agents, the performance and intra-agent dynamics (PLV and SD(KOP)) depend on the stimulus sensitivity $S$ and the internal coupling $a_{v_i,,v_j}$ of the agent. The 50 different internal coupling values in the plots correspond to $a_{v_i,v_j} \in \{0, \ldots ,2.5\}$ and the 10 values of stimulus sensitivity correspond to $c \in \{0, \ldots ,10\}$ in Eq.~\ref{eq:HKBsocial} }
\label{figure4}
\end{figure}

\section{Individual gradient ascent and binary decision-making performance}
Also in single-agent experiments, we analyze the effect of varying stimulus sensitivity and internal coupling on agent performance in gradient ascent and binary decision-making. We also analyze how agent performance is associated with intra-agent dynamics. 
A first observation is that agents without any coupling between motor oscillators (Fig.~\ref{figure4}A-I) have, in general, lower performances and a lower degree of metastable dynamics. Adding motor coupling (Fig. \ref{figure4}D-L) increases performance and metastability in the agent's intra-agent dynamics. It might seem counterintuitive that an extra coupling link in the agent's architecture leads to lower functional integration of the oscillator dynamics (indicated by lower PLV in E than B, in Fig.~\ref{figure4}). Note however, that adding a link between the motor oscillators allows the agent to have a motor response to environmental stimuli that is coordinated between the left and right sensorimotor connections.

In general, the analysis in Fig.~\ref{figure4} shows that the best performance occurs for agents that have a balance of stimulus sensitivity and internal coupling. A high degree of stimulus sensitivity without high internal coupling results in a hypersensitive agent that cannot maintain a relatively stable movement direction. Conversely, high internal coupling without high stimulus sensitivity results in an agent that cannot adequately incorporate sensory information in it's internal dynamics, which impedes changing movement direction in response to sensory input.

In our agents, sensory input does not directly control motor activity. Sensory information (in the form of stimulus concentration) is first integrated into the oscillator phase of two sensory nodes. These sensory nodes are dynamically connected to two motor nodes. 

\section{The nonuniform effect of social influence on collective decision-making}
The ability of agents to move towards one of the stimulus sources depends on their interactions with the environment: the sensory input related to the stimuli needs to have sufficient influence on the intra-agent dynamics of the agent. The agent also processes social stimuli, which should be integrated with environmental stimuli. If the social stimulation becomes too strong, it can nullify the processing of the environmental stimuli. Indeed, along the right edge of Fig.~5A in the main paper, we see that increased social influence does not uniformly increase performance; beyond some threshold, it inhibits agents' ability to reach a consensus.

\begin{figure}
\centering
\includegraphics[width=\linewidth]{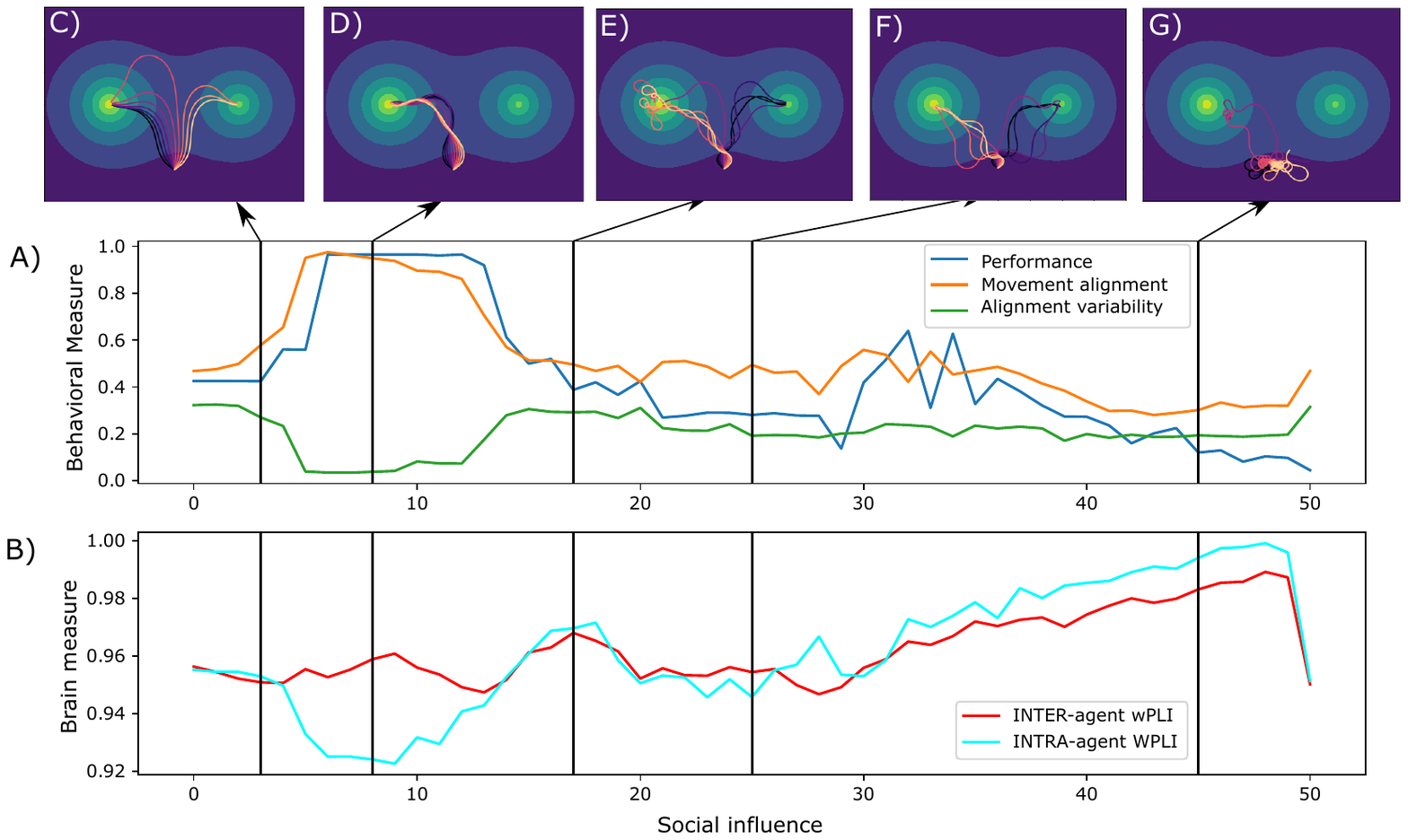} 
\caption{The effect of social influence on decision-making performance. (A) Performance, average alignment (mean KOP) and alignment variability (SD(KOP)), as a function of the degree of social influence (social sensitivity $S$ of agents). (B) intra- and inter-agent covariance of neural dynamics (wPLI). (C-G) Trajectories of collective decisions taken with increasing social influence.}
\label{figure6}
\end{figure}

Fig.~\ref{figure6}A shows a cross-section of the results shown in Fig.~5A. Increased social influence initially increases the decision-making performance (compare Fig.~\ref{figure6}C and Fig.~\ref{figure6}D), but with further increased social influence, the effect is strong enough that agents initially overshoot, in the direction that the most social influence was sensed (Fig.~\ref{figure6}E). Due to this overshooting, agents split into two groups and end up beyond the range where they can effectively influence the other group's direction.
When social influence is increased even further (Fig.~\ref{figure6}G), the effect is so strong that agents' actions are dependent only on each other, not on the environment, and most agents get into a deadlock and cannot approach a stimulus source. 

The neural measures for this cross-section (see Fig.~\ref{figure6}B) show that intra-agent wPLI decreases with respect to inter-agent wPLI in the range with the highest performance. This indicates that agents achieved consensus when their intrinsic dynamics were perturbed (i.e., uncoupled) by stimulus but remained similar (i.e., coupled) across agents. With sensitivity to social influence increases beyond this range (while internal coupling and environmental sensitivity decrease), both intra- and inter-agent wPLI increase, as social influence starts to drive each agent’s internal dynamics.

\section{Nonlinear decrease in performance due to environmental variation}

\begin{figure}
\centering
\includegraphics[width=\linewidth]{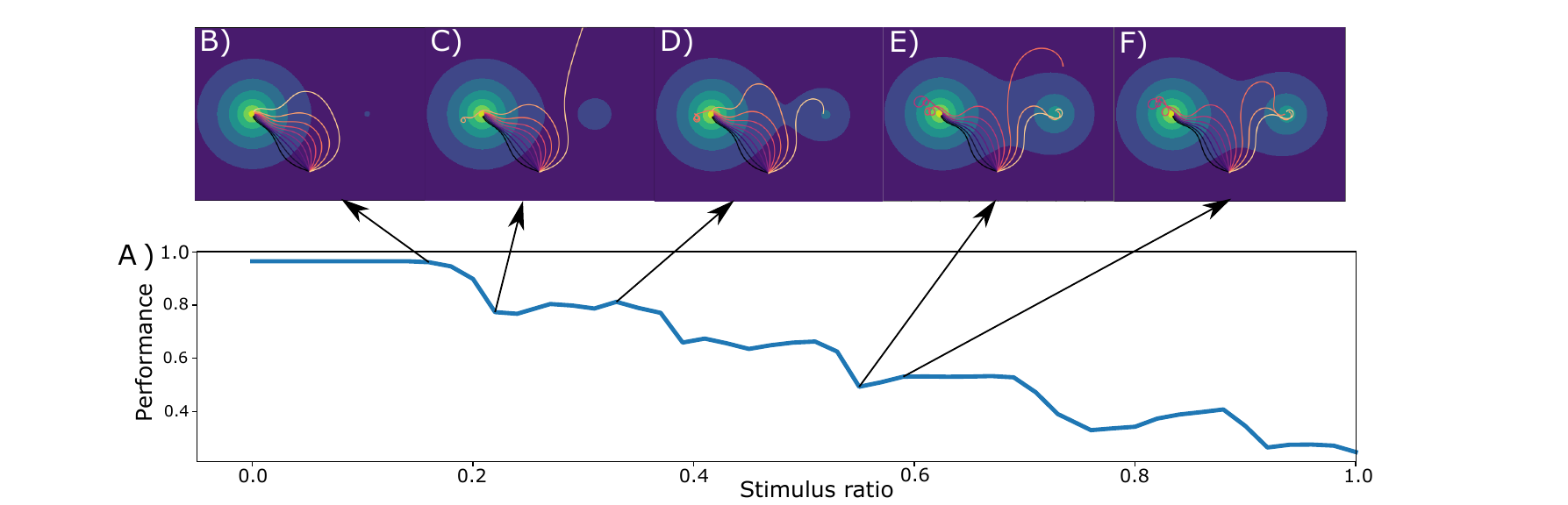} 
\caption{Decision-making performance decrease with increased stimulus ratio. (A) Performance data from Fig.~6, for the 140° starting angle. (B-F) Agent trajectories for different stimulus ratios.}
\label{figure8}
\end{figure}

To explain the apparent `plateaus' in performance in Fig.~6 of the main paper, Fig.~\ref{figure8} shows the results from one horizontal row of data in Fig.~6 
When only the left stimulus source is present (i.e., stimulus ratio 0), all agents converge to it (Fig.~\ref{figure8}B). When the stimulus source on the right attains a certain level of brightness (Fig.~\ref{figure8}C), one agent is sufficiently attracted to it to be diverted from the group and 
`catapulted' far away, 
causing a sharp performance decrease. This continues as the right source gets brighter, causing a rough plateau. At some threshold along the plateau (Fig.~\ref{figure8}D), the right source is bright enough to attract the diverted agent to its center, which results in a very slight performance bump (since the agent is also closer to the left stimulus, as a side effect). As the right stimulus becomes brighter, more agents are attracted to it, each additional agent causing a sharp decrease in performance (e.g., Fig.~\ref{figure8}E) followed by a short plateau (Fig.~\ref{figure8}F).

\section{Qualitative description of unexpected behaviors}
When some agents are individually able to reach a stimulus source, our model also shows how the spatial and time-dependent nature of the decision-making process can sometimes impede groups with higher social sensitivity from reaching a consensus. For example, in Fig.~\ref{figure6}E, agents start out with different movement directions and perceive high social information relative to stimulus information, which changes their travel directions to face towards the others. If all agents do this simultaneously and strongly, agents can \textit{overshoot}, arriving where other agents previously were, but have since left. By the time agents react to the new conditions, they can have moved so close to the different stimulus sources that they are too far apart to socially influence each other and realign towards a consensus. 
If social influence were less strong, the initial overshoot would be smaller, allowing agents to remain closer together and better balance social and environmental information (as in Fig.~\ref{figure6}d). In extreme cases, some agents might experience a combination of social influence that leaves them moving away from the group and undecided for too long, so that they have moved past the window in which stimulus information was salient and moved further away from both stimulus sources, illustrating how certain patterns of social influence can lead some individuals to isolation.

\bibliographystyle{plain} 
\bibliography{references} 